\documentclass[onecolumn,nofootinbib,superscriptaddress,notitlepage,amsmath,amssymb,fleqn,aps,prd,preprintnumbers,10pt]{revtex4-2}
\usepackage[table]{xcolor}
\usepackage{graphicx,xspace,hyperref,cancel}
\usepackage[all]{hypcap}
\usepackage{todonotes}
\usepackage{booktabs,diagbox,multirow}
\usepackage{siunitx}
\newcommand{\eps}{\varepsilon}

\hypersetup{hidelinks,
  pdfauthor={John M. Campbell,Stefan Hoeche,Max Knobbe},
  pdftitle={Double-real corrections to e+e- to qq with partial fractioned scalar radiators and identified particle kinematics mappings},
  pdfkeywords={QCD, Perturbation Theory}
}

\begin{document}
\preprint{FERMILAB-PUB-26-0308-T, MCNET-26-20}
\title{Double-real corrections to \texorpdfstring{\boldmath{$e^+e^-\to q\bar{q}$}}{e+e- to qq}
  with partial fractioned\\ scalar radiators and identified particle kinematics mappings}
\author{John M.\ Campbell}
\affiliation{Fermi National Accelerator Laboratory, Batavia, IL, 60510, USA}
\author{Stefan H{\"o}che}
\affiliation{Fermi National Accelerator Laboratory, Batavia, IL, 60510, USA}
\author{Max Knobbe}
\affiliation{Fermi National Accelerator Laboratory, Batavia, IL, 60510, USA}

\begin{abstract}
    We present a new approach to local infrared subtraction for next-to-next-to-leading
    order QCD calculations in $e^+e^-\to q\bar{q}$. Overlapping singularities in the
    multipole radiation pattern are disentangled by partial fractioning, and the kinematics
    mapping corresponds to iterated next-to-leading order identified particle kinematics.
    We verify that the double-real remainder is rendered finite in the single and
    double unresolved limits and investigate the numerical convergence of the
    Monte-Carlo integral. We compute the phase-space integrals of the scalar counterterms
    in the back-to-back configuration, both analytically and with the help of
    numerical techniques based on sector decomposition.
\end{abstract}

\maketitle

\section{Introduction}
Calculations based on Quantum Chromodynamics (QCD) have been extraordinarily successful
in predicting the structure of events observed in high-energy physics experiments
that started with the first electron-positron colliders about fifty years
ago~\cite{Gross:2022hyw}. While theory continues to keep pace with experiment in general,
the outstanding performance of the Large Hadron Collider (LHC), and the high precision
of measurements made by the ATLAS, CMS, LHCb and ALICE collaborations have highlighted
the need for many more high-precision calculations than might have previously been
anticipated~\cite{Huss:2022ful,Huss:2025nlt}. The developing gap in precision could widen
dramatically at an electron-positron Future Circular Collider (FCC-ee), which is currently
discussed as a potential successor to the LHC at CERN~\cite{FCC:2025jtd,FCC:2025uan,FCC:2025lpp}.

The fully differential higher-order QCD calculations needed to close this gap are
complicated by soft-gluon divergences and collinear singularities, which are guaranteed
to cancel between real and virtual corrections by the Bloch-Nordsieck theorem~\cite{Bloch:1937pw}
and the Kinoshita-Lee-Nauenberg theorem~\cite{Kinoshita:1962ur,Lee:1964is}. In order
to perform higher-order calculations in practice, one must identify the singularities,
and isolate them so as to only compute the finite remainder which is observable.
This is usually complicated by the fact that the singularities overlap, even at the
first nontrivial order. Presently available techniques to address this problem
at the next-to-next-to-leading order (NNLO) in QCD include sector decomposition~\cite{
  Anastasiou:2003gr,Anastasiou:2005qj,Binoth:2004jv,Melnikov:2006kv},
antenna subtraction~\cite{Gehrmann-DeRidder:2005alt,
  Gehrmann-DeRidder:2005btv,Daleo:2006xa,Daleo:2009yj,Gehrmann-DeRidder:2012too,Currie:2013vh},
the CoLoRFulNNLO method~\cite{Somogyi:2005xz,Somogyi:2006da,DelDuca:2016ily,DelDuca:2024ovc,DelDuca:2025yph},
sector-improved residue subtraction~\cite{Czakon:2010td,Czakon:2011ve,Boughezal:2011jf,Czakon:2014oma},
nested soft–collinear subtraction~\cite{Caola:2017dug,Devoto:2025kin,Devoto:2025jql},
local analytic sector subtraction~\cite{Magnea:2018ebr,Magnea:2018hab,Magnea:2020trj},
projection-to-Born~\cite{Cacciari:2015jma,Campbell:2024hjq} and slicing-based
approaches~\cite{Catani:2007vq,Boughezal:2015eha,Boughezal:2015dva,Gaunt:2015pea}.
These methods differ in the precise form of the infrared counterterms,
and in how the counterterms are used to cover the singular and regular regions
of the phase space of the multi-parton final state.
 
In this work we present a new calculation of the NNLO double-real corrections in $e^+e^-\to q\bar{q}$,
that builds on a recently introduced algorithm for assembling splitting functions from spin-independent
and spin-dependent components~\cite{Campbell:2025lrs,Hoche:2025vto}. The main benefit of the technique
is that it provides a straightforward separation of singular structures, and that there is a
correspondence to resummed calculations in the form of parton showers~\cite{Campbell:2022qmc}.
The kinematic mappings in the double-real counterterms are simple iterations of the mappings
in the single-real case, with a parametrization that matches the recoil definition of a
number of NLL-safe parton-shower algorithms~\cite{Herren:2022jej,Assi:2023rbu,
  Hoche:2025anb,Hoche:2026dup,Preuss:2024vyu}.

In order to introduce a practical NNLO subtraction it is important to demonstrate
that the double-real subtraction terms can be integrated over the unresolved phase-space
to expose poles in a form suitable for cancellation against other elements of the calculation.
Here we show that this is the case in the back-to-back kinematic configuration.    
The phase-space integrals of the scalar radiators can be computed in this particular case
using reverse unitarity~\cite{Gehrmann-DeRidder:2003pne}, utilizing Kira~\cite{Maierhofer:2017gsa,
  Lange:2025fba} to reduce all integrals to a known set of masters. We also verify
that a numerical approach based on Laurent series expansion and
sectorization~\cite{Czakon:2010td,Czakon:2011ve} can alternatively be used. 
  
We implement the double-real counterterms in the
\textsc{Pepper}~\cite{Bothmann:2021nch,Bothmann:2023siu,Bothmann:2023gew}
and \textsc{Sherpa}~\cite{Gleisberg:2008ta,Sherpa:2019gpd,Sherpa:2024mfk} event generators
and validate them in all kinematic limits relevant to hadronic final state production
at a lepton collider. The algorithms discussed in Ref.~\cite{Bothmann:2024hgq}
allow us to probe the double-real remainder in all
single- and double-unresolved regions and demonstrate the expected
cancellation of singularities in the deep infrared. Finally, we numerically integrate
the subtracted double-real emission contribution to $e^+e^-\to q\bar{q}$ and present some
preliminary distributions. These results by themselves are unphysical, as they lack the real,
real-virtual and double virtual corrections. However, they provide an estimate of the
expected numerical performance of the complete infrared subtraction in the setting of
a Tera-$Z$ option of the FCC-ee, where not only high parametric precision,
but also extreme statistical accuracy is required.

The manuscript is structured as follows. Section~\ref{sec:method} explains the
method of calculation. We focus on the double-real corrections, because they are
typically the most complicated numerically.
The validation of the calculation and an assessment of its performance are presented
in Sec.~\ref{sec:numerics}. Section~\ref{sec:outlook} contains an outlook.

\section{The method}
\label{sec:method}
In this section we discuss the formulation of the double-real subtraction terms. The
required splitting functions have been derived in Ref.~\cite{Campbell:2025lrs}.
We will not list them again, except when necessary for the understanding of the method.
The main challenge addressed in this manuscript is the combination of
scalar radiators and pure splitting functions, as well as the combination of the NNLO
and NLO subtraction terms, which is needed to cancel singularities in both the
single- and double-unresolved limits.

\subsection{Structure of the result}
\label{sec:structure}
The general aspects of combining NLO and NNLO counterterms have been
discussed in Ref.~\cite{Gehrmann-DeRidder:2004ttg}, and we refer the reader
to the original publication for an introduction to the topic. Schematically,
we can write the complete result of a subtracted double-real squared matrix element
in a process with $n$ partons at Born level in the following form
\begin{equation}\label{eq:structure}
  \mathcal{M}_{RR} = \mathcal{M}_{n+2} -\sum \mathcal{M}_n\otimes S_2
  -\sum \Big[\,\mathcal{M}_{n+1} - \mathcal{M}_n\otimes S_1\,\Big]\otimes S_1\;.
\end{equation}
Here, $\mathcal{M}_n$ denotes the squared tree-level matrix element
with $n$ final-state particles, $S_1$ denotes the NLO subtraction terms,
$S_2$ denotes the NNLO subtraction terms, and $\otimes$ indicates that
there are possible color and spin correlations between the individual
components. For ease of notation, we suppress the coupling constant
here and in the following. In color singlet production, scattering or decay,
color and kinematics factorize, which simplifies the calculation considerably.
While the terms involving $\mathcal{M}_n\otimes S_2$ remove all
divergences in the double-unresolved regions, the remaining terms
in Eq.~\eqref{eq:structure} are required to cancel the singularities
of $\mathcal{M}_{n+2}$ in the single-unresolved limits, but without
double-counting any of the double-unresolved configurations already
addressed by $\mathcal{M}_n\otimes S_2$~\cite{Gehrmann-DeRidder:2004ttg}.
Equation~\eqref{eq:structure} can alternatively be written as
\begin{equation}\label{eq:structure_2}
  \mathcal{M}_{RR} = \mathcal{M}_{n+2}-\sum \mathcal{M}_n\otimes S_1\otimes S_1
  -\sum \mathcal{M}_n\otimes\Big[\,S_2-\sum S_1\otimes S_1\,\Big]
  -\sum \Big[\,\mathcal{M}_{n+1} - \sum\mathcal{M}_n\otimes S_1\,\Big]\otimes S_1\;,
\end{equation}
which emphasizes that the complete set of counterterms consists of
the strongly ordered terms of squared NLO type, $S_1\otimes S_1$, the pure
NNLO remainders, $S_2-\sum S_1\,\otimes S_1$, and remainders that contain
the product of the infrared finite $n+1$-parton contribution $\mathcal{M}_{n+1}
  -\sum\mathcal{M}_n\otimes S_1$ with NLO-type subtraction terms, $S_1$.
The first subtracted term contributes in all unresolved regions, while the second
and third contribute only in the double- and single-unresolved regions,
respectively. We note that a parton shower based on spin- and color-correlated
leading-order splitting kernels will generate precisely the first subtraction
term in Eq.~\eqref{eq:structure_2}~\cite{Hoeche:2011fd,Hoche:2017iem,Dulat:2018vuy}.

\subsection{Kinematics and choice of gauge vector}
\label{sec:preliminaries}
Beyond the color and spin correlations, the convolution symbol $\otimes$
in Eq.~\eqref{eq:structure_2} indicates that terms to the left of it should be
evaluated with on-shell parton momenta. This is achieved by means of a momentum
mapping, which becomes an essential component of the infrared subtraction. 
We aim at a method that will eventually allow a matching to parton showers
with NLL-preserving kinematics~\cite{Herren:2022jej,Hoche:2024dee,
Hoche:2025gsb}, and therefore we rely on the momentum mapping
in Sec.~5.6 of Ref.~\cite{Catani:1996vz}. The details of this algorithm
and the related phase-space factorization are discussed in
App.~\ref{sec:ps_factorization}. One starts with an identified momentum,
$\tilde{p}_i^\mu$, and a recoil momentum, $\tilde{K}^\mu$.
The recoil momentum can be defined such as to reflect the physics of the process
under consideration. The mapping must be collinear safe and conserve four-momentum,
two constraints that are satisfied by the definition
\begin{equation}\label{eq:def_n_pi}
    p_i^\mu=z\,\tilde{p}_{i}^\mu\;,\qquad
    N^\mu=\tilde{K}^\mu+(1-z)\,\tilde{p}_{i}^\mu\;,
\end{equation}
where $z=(p_iN)/(\tilde{p}_iN)$ is the identified parton's forward momentum fraction.
To generate the momentum of the emission, $q_1^\mu$, we set $N^\mu=K^\mu+q_1^\mu$
and use a Sudakov decomposition~\cite{Sudakov:1954sw} along $\tilde{p}_i^\mu$
and $N^\mu$. To perform the phase-space integrals at NNLO, we need to apply this
kinematics mapping twice. Let us assume that the vector $\tilde{p}_i$ absorbs the
longitudinal recoil of both emissions. Applying Eq.~\eqref{eq:def_n_pi} twice,
we then obtain
\begin{equation}\label{eq:final_recoil_two_emissions}
  K^\mu+q_1^\mu+q_2^\mu=\tilde{K}^\mu+(1-z)(1-z')\,\tilde{p}_i^\mu\;,
\end{equation}
with suitably defined momentum fractions, $z$ and $z'$.
The result is symmetric in the two emissions and independent of the 
splitting sequence, which also applies if the second recoil momentum
is $q_1$. This allows us to parametrize the most involved differential
phase-space elements needed for the integrals in Sec.~\ref{sec:overlap_removal}
in a symmetric fashion\footnote{
  Details on the phase-space parametrization are given in
  App.~\ref{sec:ps_factorization}.}.

In order to evaluate the pure splitting functions from Ref.~\cite{Campbell:2025lrs},
we need to define a gauge vector to perform the Sudakov decomposition~\cite{Sudakov:1954sw}.
To simplify the calculation, such vectors are typically chosen to be light-like.
For two momenta, $p_i^\mu$ and $p_j^\mu$, one can then write
\begin{equation}\label{eq:sud_simple}
  \begin{split}
    p_i^\mu=&\;z_{i,j}\, p_{ij}^\mu
    -\bar{p}_{i,j}^\mu\;,
    \qquad\text{and}\qquad
    &p_j^\mu=&\;z_{j,i}\, p_{ij}^\mu
    -\bar{p}_{j,i}^\mu\;,
  \end{split}
\end{equation}
where $p_{ij}^\mu=p_i^\mu+p_j^\mu$.
The individual components are
\begin{equation}\label{eq:def_z_ptilde_1}
    z_{i,j}=\frac{p_i\bar{n}}{p_{ij}\bar{n}}\;,
    \qquad\text{and}\qquad
   \bar{p}_{i,j}^\mu=\frac{z_{i,j}p_j^\mu-z_{j,i}p_i^\mu}{z_{i,j}+z_{j,i}}\;.
\end{equation}
In the present manuscript we will instead use a time-like gauge vector,
which will conveniently be identified with the vector $N^\mu$ from
Eq.~\eqref{eq:def_n_pi}. This corresponds to a choice of reference frame
in which scaled energies and angular variables are defined as
\begin{equation}\label{eq:def_eta}
    \zeta_i=\frac{2p_in}{n^2}\;,
    \qquad\text{and}\qquad
    \eta_{ij}=\frac{2p_ip_j}{\zeta_i\zeta_j\,n^2}\;. 
\end{equation}
When the two momenta in Eq.~\eqref{eq:sud_simple} become collinear,
we have $p_i\bar{n}=p_in+\mathcal{O}(p_{ij}^2)$, such that
-- to leading power in the collinear scaling parameter --
the light-like axial gauge defined by $\bar{n}^\mu$ yields
the same splitting functions as the time-like axial gauge
defined by $n^\mu$. In particular, the expressions from
Ref.~\cite{Campbell:2025lrs} remain valid at leading power,
and we can compute them in the following parametrization:
\begin{equation}\label{eq:def_z_ptilde}
    p_i^\mu=\frac{\zeta_i}{\zeta_i+\zeta_j}\, p_{ij}^\mu
    -\tilde{p}_{i,j}^\mu\;,
    \qquad\text{where}\qquad
   \tilde{p}_{i,j}^\mu=\frac{\zeta_ip_j^\mu-\zeta_jp_i^\mu}{\zeta_i+\zeta_j}\;.
\end{equation}
The scalar radiators will instead be rearranged into gauge-invariant
subsets that match the typical definition in a parton shower.
This is discussed in the next section.

\subsection{Overlapping singularities in scalar radiators}
\label{sec:overlap_removal}
In this section, we describe how soft-enhanced components of the hard
matrix elements are disentangled such that singularities can be attributed
to unique collinear configurations. This is required to satisfy collinear
safety, which has also been built into the kinematics mapping introduced
in Sec.~\ref{sec:preliminaries}. Only after the collinear singularities
have been made explicit is it possible to apply the phase-space parametrizations
in App.~\ref{sec:ps_factorization} to extract the infrared poles.
The identification is hampered mostly by the fact
that gluons are radiated coherently off QCD charge multipoles.
While it is not possible to attribute their production to individual charges,
one could use an axial gauge to separate the complete multipole radiator
into collinear splitting functions and wide-angle soft remainders~\cite{
  Marchesini:1987cf,Marchesini:1989yk}.
At this point it is important to recall that we attempt to construct
an infrared subtraction which may allow a matching to parton showers.
The choice of an axial gauge corresponds to an additive matching of radiators,
which is rather impractical because it creates splitting functions that assume
negative values in large regions of the phase space and thus require
azimuthal averaging~\cite{Herren:2022jej}.
A better approach is to partial fraction the sum of scalar radiators,
using the methods introduced in Refs.~\cite{Ellis:1980wv,Catani:1996vz}.
The following two subsections will explain how this can be achieved
at NNLO.

\subsubsection{The NLO case}
\label{sec:overlap_removal_nlo}
We begin the discussion with the simple example of an NLO calculation.
The scalar radiator for the production of a single on-shell gluon
from a QCD multipole formed by massless on-shell particles is given
by~\cite{Bassetto:1984ik,Campbell:2025lrs}
\begin{equation}\label{eq:one-gluon_radiator}
  \begin{split}
  \mathcal{S}_g(\{p\};q_1) = &\; \sum_{i,k}
    \hat{\bf T}_i\hat{\bf T}_k\,
    \mathcal{S}_{i;k}(q_1)\;,
  \qquad\text{where}\qquad
  \mathcal{S}_{i;k}(q_1)=-\frac{4p_ip_k}{p_{i1}^2p_{k1}^2}\;.
  \end{split}
\end{equation}
Note that we have used current conservation to eliminate the gauge
dependent terms\footnote{The gauge dependence in the off-shell
  effects on Eq.~(46) of Ref.~\cite{Campbell:2025lrs} cannot be
  eliminated. Therefore, any off-shell scalar radiators multiplying
  purely fermionic remainder functions must still be evaluated
  in axial gauge. This applies in particular
  to Eqs.~(70) and~(80) of Ref.~\cite{Campbell:2025lrs}.}.
We must disentangle the overlapping singularities in Eq.~\eqref{eq:one-gluon_radiator}
to avoid soft double-counting when the infrared counterterms are used
as splitting kernels in a parton shower. To this end, we employ
techniques from Refs.~\cite{Herren:2022jej,Assi:2023rbu}.
Using the angular variable in Eq.~\eqref{eq:def_eta},
we partial fraction the radiator as follows
\begin{equation}\label{eq:one-gluon_radiator_part}
    \mathcal{S}_{i;k}(q_1)=
    \mathcal{S}_{i;k}^{(i)}(q_1;n)+
    (i\leftrightarrow k)\;,
    \qquad\text{where}\qquad
    \mathcal{S}_{i;k}^{(i)}(q_1;n)=
    \frac{\eta_{k1}}{\eta_{k1}+\eta_{i1}}\,
    \mathcal{S}_{i;k}(q_1)\;.
\end{equation}
At next-to-leading order in QCD the only overlapping singularities
appear in the eikonal factors, $\mathcal{S}_{i;k}$.
Equation~\eqref{eq:one-gluon_radiator_part} is therefore sufficient
to construct subtraction terms with a unique assignment of the parton
that plays the role of the emitter in the momentum mapping algorithm
of Sec.~\ref{sec:preliminaries}. The phase-space parametrization
discussed in App.~\ref{sec:ps_factorization_rad_fi} is sufficient to
perform the relevant integrals, the corresponding results can be found
in Ref.~\cite{Hoche:2025gsb}.

\subsubsection{The NNLO case -- Abelian contributions}
\label{sec:overlap_removal_nnlo_ab}
References~\cite{Catani:1999ss,Campbell:2025lrs} discussed the
process-independent form of the two-gluon soft / scalar radiators,
which can be determined from the squared two-gluon current.
The abelian component of the scalar radiator is given by
\begin{equation}\label{eq:squared_two-gluon_current_ab}
  \begin{split}
    &S_{gg}^{(\mathrm{ab})}(\{p\},q_1,q_2)=2\sum_{i,k}\sum_{l,m}
    \Big\{\hat{\bf T}^a_i\hat{\bf T}^a_l,
    \hat{\bf T}^b_k\hat{\bf T}^b_m\Big\}\,
    \mathcal{S}^{\rm(ab)}_{i,k;l,m}(q_1,q_2)\\
    &\;\quad\quad+2\sum_{i,k}\sum_{l}
    \Big(\Big\{\hat{\bf T}^a_i\hat{\bf T}^a_l,
    \hat{\bf T}^b_k\hat{\bf T}^b_l\Big\}+
    \Big\{\hat{\bf T}^a_l\hat{\bf T}^a_i,
    \hat{\bf T}^b_l\hat{\bf T}^b_k\Big\}\Big)\,
    \mathcal{S}^{\rm(ab)}_{i,k;l}(q_1,q_2)
    +2\sum_{i,l}\Big\{\hat{\bf T}^a_i\hat{\bf T}^a_l,
    \hat{\bf T}^b_i\hat{\bf T}^b_l\Big\}\,
    \mathcal{S}^{\rm(ab)}_{i;l}(q_1,q_2)\,.
    \end{split}
\end{equation}
Again, we can use current conservation to rearrange the terms
in the sum such that the rather involved expressions in
Ref.~\cite{Campbell:2025lrs} are simplified. The resulting,
gauge-independent radiators read
\begin{equation}\label{eq:simplified_abelian_radiators}
    \begin{split}
    \mathcal{S}^{\rm(ab)}_{i,k;l,m}(q_1,q_2)=&\;
    \frac{1}{4}\,\mathcal{S}_{i;l}(q_1)\mathcal{S}_{k;m}(q_2)\;,\\
    \mathcal{S}^{\rm(ab)}_{i,k;l}(q_1,q_2)=&\;
    \frac{1}{s_{l12}}\frac{s_{il}s_{kl}}{s_{i1}s_{k2}}
    \bigg(\frac{s_{k1}}{s_{kl}s_{l1}}+\frac{s_{i2}}{s_{il}s_{l2}}
    -\frac{s_{ik}}{s_{il}s_{kl}}-\frac{s_{12}}{s_{l1}s_{l2}}\bigg)\;,\\
    \mathcal{S}^{\rm(ab)}_{i;l}(q_1,q_2)=&\;
    \frac{\lambda^2(s_{i1}s_{l2},s_{i2}s_{l1},s_{12}s_{il})}{
    s_{i12}s_{l12}\,s_{i1}s_{l1}s_{i2}s_{l2}}
    +\frac{2(1-\eps)}{s_{i12}s_{l12}}\;,
    \end{split}
\end{equation}
where $\lambda^2(a,b,c)=(a-b-c)^2-4bc$, and where $s_{ik}=2p_ip_k$.
It is interesting to note that the sub-leading contributions,
$\mathcal{S}^{\rm(ab)}_{i,k;l}$, have only integrable singularities
associated with any double-collinear limit, they describe purely soft and
triple-collinear radiative effects. This is made explicit by writing them as
\begin{equation}\label{eq:sikl_azimuthal_correlation}
    \mathcal{S}^{\rm(ab)}_{i,k;l}(q_1,q_2)=
    \frac{2\cos\phi_{1i,k2}^{(l)}}{s_{l12}}\sqrt{\frac{s_{il}}{s_{i1}s_{l1}}
    \frac{s_{kl}}{s_{k2}s_{l2}}}\;,
\end{equation}
where the azimuthal angle is given in terms of linear polarization
vectors along $p_l^\mu$~\cite{Dulat:2018vuy,Hoche:2020pxj}:
\begin{equation}\label{eq:azimuthal_correlation_polvec}
    \cos\phi_{1i,k2}^{(l)}=\epsilon_{1i,l}^\mu\epsilon_{k2,l\,\mu}\;,
    \qquad\text{where}\qquad
    \epsilon_{ik,l}^\mu=\frac{(p_lp_i)p_k^\mu-(p_lp_k)p_i^\mu}{
     \sqrt{2(p_ip_k)(p_lp_i)(p_lp_k)}}\;.
\end{equation}
The sub-sub-leading terms, $\mathcal{S}^{\rm(ab)}_{i;l}$,
can be written as an azimuthal correlation between $p_1^\mu$ and
$p_2^\mu$ in the frame where $p_i^\mu$ and $p_l^\mu$ are aligned
along the positive and negative $z$-axis, the $il$-dipole rest frame.
 \begin{equation}\label{eq:sil_azimuthal_correlation}
    \begin{split}
    \mathcal{S}^{\rm(ab)}_{i;l}(q_1,q_2)=&\;
    \frac{2(1-\eps)-4\sin^2\phi_{1,2}^{\,i,l}}{s_{i12}s_{l12}}\;.
    \end{split}
\end{equation}
The contribution proportional to $\sin^2\phi_{1,2}^{\,i,l}$
only produces integrable singularities, and the one proportional
to $1-\eps$ remains integrable when $i\neq l$ (see also Eq.~(4.25)
of Ref.~\cite{Gehrmann-DeRidder:2003pne}). Some of the terms are
therefore not needed in the subtraction procedure, but we decide
to retain them, as they present no additional complication and will
likely improve numerical convergence.

We perform a partial fractioning of the form of
Eq.~\eqref{eq:one-gluon_radiator_part} on the individual NLO-like
terms in $\mathcal{S}^{\rm(ab)}_{i,k;l,m}(q_1,q_2)$:
\begin{equation}\label{eq:two-gluon_radiator_ab_part_1x}
    \mathcal{S}^{{\rm(ab)},(i,k)}_{i,k;l,m}(q_1,q_2)=
  \frac{1}{4}\,\mathcal{S}_{i;l}^{(i)}(q_1;K+q_{12})\,
  \mathcal{S}_{k;m}^{(k)}(q_2;K+q_{12})\;,
\end{equation}
with the individual factors given by Eq.~\eqref{eq:one-gluon_radiator_part}.

A crucial question is whether the phase-space integrals of the above
subtraction terms can be obtained. Here we compute them in the very simple
but phenomenologically important setting of a back-to-back Born configuration.
The restricted kinematics allows a recombination of the partial fractioned radiators
at the integrand level. We can then use reverse unitarity, as described in
Ref.~\cite{Gehrmann-DeRidder:2003pne}, with the master integrals taken
from the same reference. The integral reduction was performed using
Kira~\cite{Maierhofer:2017gsa,Lange:2025fba}.
Defining the coefficient functions
\begin{equation}\label{eq:def_integrals}
  {\cal I}_X = \frac{\big(\tilde{K}^2\big)^{2\eps}}{4}
  \left[ \frac{e^{\eps\gamma_E}}{\Gamma(1-\eps)} \right]^2
  \frac{1}{S_\Gamma}\int {\mathrm d}PS_4 \, {\cal S}_X\;,
  \qquad\text{where}\qquad
  S_\Gamma=P_2\bigg(\frac{(4\pi)^\eps}{16\pi^2\,\Gamma(1-\eps)}\bigg)^2\;,
\end{equation}
in $D=4-2\eps$ dimensions, with $P_2=2^{-3+2\eps}\pi^{-1+\eps}\Gamma(1-\eps)/\Gamma(2-2\eps)
  \big(\tilde{K}^2\big)^{-\eps}$, we find (systematically dropping
${\cal O}(\epsilon)$ terms, and assuming $i\neq k$)
\begin{eqnarray}\label{eq:integral_Sab_iikk}
  {\cal I}_{i,i;k,k}^{\rm(ab)}&=&
 \frac{1}{4\epsilon^4}+\frac{1}{\epsilon^3}+\frac{1}{\epsilon^2}\left(\frac{9}{2}-\frac{3}{8}\pi^2\right)
 +\frac{1}{\epsilon}\left(19-\frac{37}{6}\zeta_3-\frac{3}{2}\pi^2\right)
 +\left(78-\frac{77}{3}\zeta_3-\frac{27}{4}\pi^2+\frac{49}{480}\pi^4\right)\;,\\
  \label{eq:integral_Sab_iik}
  {\cal I}_{i,i;k}^{\rm(ab)}&=&
  \frac{1}{2\epsilon}\Big(1-\zeta_3\Big)+\left(\frac{9}{2}-\zeta_3-\frac{\pi^4}{24}\right)\;,\\
  \label{eq:integral_Sab_ik}
  {\cal I}_{i;i}^{\rm(ab)}&=&-\frac{1}{8\epsilon}-\frac{15}{16}\;,
  \qquad\text{and}\qquad
  {\cal I}_{i;k}^{\rm(ab)}=-\frac{1}{2}+\frac{1}{12} \pi^2-\frac{1}{360} \pi^4\;.
\end{eqnarray}
Note that ${\cal I}_{i,k;k,i}^{\rm(ab)} = {\cal I}_{i,i;k,k}^{\rm(ab)}$ and ${\cal I}_{i,k;i}^{(ab)}={\cal I}_{i,k;k}^{\rm(ab)}=0$,
since the integrands vanish.
In App.~\ref{sec:ps_factorization_rad_fi_nnlo} we discuss a phase-space
parametrization which, when combined with the sector decomposition
from Refs.~\cite{Czakon:2010td,Czakon:2011ve}, allows us to compute
Eqs.~\eqref{eq:integral_Sab_iikk}-\eqref{eq:integral_Sab_ik} through
numerical integration of the partial fractioned scalar radiators.
Results from this method are presented in Sec.~\ref{sec:numerics}.

\subsubsection{The NNLO case -- Nonabelian contributions}
\label{sec:overlap_removal_nnlo_nab}
The non-abelian scalar radiator was computed in Ref.~\cite{Campbell:2025lrs}.
As in the abelian case, one can use charge conservation to simplify the expressions
considerably. We obtain
\begin{equation}\label{eq:squared_two-gluon_current_nab}
  \begin{split}
    S_{gg}^{(\mathrm{nab})}(\{p\},q_1,q_2)=
    \sum_{i,l}C_A\hat{\bf T}^c_i\hat{\bf T}^c_l\,
    \Big[\,\mathcal{S}^{\rm(ab)}_{i,i;l}(q_1,q_2)
    +\mathcal{S}^{\rm(ab)}_{l,l;i}(q_1,q_2)
    +(1-2\delta_{il})\,\mathcal{S}^{\rm(ab)}_{i;l}(q_1,q_2)
    -\mathcal{S}^{\rm(nab)}_{i;l}(q_1,q_2)\,\Big]\;.
  \end{split}
\end{equation}
Note in particular that terms of the form $\mathcal{S}^{\rm(ab)}_{i,l;l}$
vanish due to Eq.~\eqref{eq:simplified_abelian_radiators}. The result for
the non-abelian dipole is
\begin{equation}\label{eq:two-gluon_radiator_nab}
  \begin{split}
    &\mathcal{S}^{\rm(nab)}_{i;k}(q_1,q_2)=
    \mathcal{S}^{\rm(nab,\,s.o.)}_{i;k}(q_1,q_2)
    -\frac{4}{s_{i12}s_{k12}}\bigg(\frac{2s_{ik}}{s_{12}}-1\bigg)
    +(1-\eps)\bigg(\frac{s_{i1}-s_{i2}}{s_{i12}s_{12}}-\frac{s_{k1}-s_{k2}}{s_{k12}s_{12}}\bigg)^2\\
    &\;\qquad-\bigg[\frac{s_{i1}s_{k2}}{s_{i12}s_{k12}}
    \bigg(1-\frac{s_{i2}+s_{k1}}{s_{ik}}-\frac{s_{12}}{2s_{ik}}\bigg)
    +\frac{s_{12}}{2s_{i12}}\bigg(1-\frac{s_{12}}{2s_{k12}}\bigg)
    +\big(i\leftrightarrow k\big)\bigg]\,
    \frac{1}{2}\,\mathcal{S}^{\rm(nab,\,s.o.)}_{i;k}(q_1,q_2)\;,
  \end{split}
\end{equation}
with the leading-power terms given by Eq.~(110) of Ref.~\cite{Catani:1999ss}
and the contribution proportional to $(1-\eps)$ exposing the structure of
polarization correlations (see Eq.~(21) in Ref.~\cite{Hoche:2025anb}).
In the strongly ordered limit, Eq.~\eqref{eq:two-gluon_radiator_nab}
reduces to
\begin{equation}\label{eq:two-gluon_radiator_nab_so}
  \begin{split}
    \mathcal{S}^{\rm(nab,\,s.o.)}_{i;k}(q_1,q_2)=&\;
    4\,\bigg(\frac{s_{ik}}{s_{i1}s_{12}s_{k2}}
    +\frac{s_{ik}}{s_{k1}s_{12}s_{i2}}
    -\frac{s_{ik}^2}{s_{i1}s_{k1}s_{i2}s_{k2}}\bigg)
    =\frac{8s_{ik}}{s_{12}}\,
    \frac{\cos\phi_{1,2}^{\,i,k}}{\sqrt{s_{i1}s_{k1}s_{i2}s_{k2}}}\;,
  \end{split}
\end{equation}
where the latter form serves to highlight the singularity structure.
Note in particular that Eq.~\eqref{eq:two-gluon_radiator_nab_so} has only
integrable double-collinear singularities except in $s_{12}$, but it does have
entangled triple-collinear poles. The mapping to the individual collinear
sectors can therefore be performed by an angular partial fractioning
of the form of Eq.~\eqref{eq:one-gluon_radiator_part}.
To simplify the numerical integration, we use an angular partition
with a straightforward structure in terms of the phase-space variables
in App.~\ref{sec:ps_factorization_rad_fi_nnlo}:
\begin{equation}\label{eq:two-gluon_radiator_nab_part_1}
    \mathcal{S}^{\rm(nab)}_{i;k}(q_1,q_2)=
    \mathcal{S}^{({\rm nab}),(i)}_{i;k}(q_1,q_2;n)+
    (i\leftrightarrow k)\;,
\end{equation}
where
\begin{equation}\label{eq:two-gluon_radiator_nab_part_eta}
    \mathcal{S}^{({\rm nab}),(i)}_{i;k}(q_1,q_2;n)=
    \frac{\eta_{k1}+\eta_{k2}}{\eta_{k1}+\eta_{k2}+\eta_{i1}+\eta_{i2}}\,
    \mathcal{S}^{\rm(nab)}_{i;k}(q_1,q_2)\;.
\end{equation}
The integrals of the above functions can be carried out in the back-to-back
configuration by first recombining their integrands. The results are,
in the notation of Eq.~\eqref{eq:def_integrals} (again assuming $i\neq k$),
\begin{eqnarray}\nonumber
  {\cal I}_{i;k}^{\rm(nab)}&=&\frac{1}{2\epsilon^4}+\frac{35}{12\epsilon^3}
  +\frac{1}{\epsilon^2}\left(\frac{523}{36}-\frac{2}{3}\pi^2\right)
  +\frac{1}{\epsilon}\left(\frac{14575}{216}-\frac{59}{6}\zeta_3-\frac{313}{72}\pi^2\right)\\
  &&\label{eq:integral_Snab_ik}
  \quad+\left(\frac{388159}{1296}-\frac{1409}{18}\zeta_3
  -\frac{4691}{216}\pi^2+\frac{47}{240}\pi^4\right)\;,\\
  \label{eq:integral_Snab_ii}
  {\cal I}_{i;i}^{\rm(nab)}&=&
  \frac{1}{\epsilon^2}+\frac{7}{\epsilon}+\bigg(37-\frac{3}{2}\pi^2\bigg)\;.
\end{eqnarray}
The result for the strongly-ordered component in
Eq.~\eqref{eq:two-gluon_radiator_nab_so} is, for $i\neq k$
\begin{equation}\label{eq:interal_Snabso_ik}
  {\cal I}_{i;k}^{\rm(nab,\,s.o.)}=
 \frac{1}{2\epsilon^4}+\frac{2}{\epsilon^3}
 +\frac{1}{\epsilon^2}\left(8-\frac{3}{4}\pi^2\right)+\frac{1}{\epsilon}
 \left(32-\frac{46}{3}\zeta_3-3\pi^2\right)
 +\left(128-\frac{172}{3}\zeta_3-12\pi^2-\frac{11}{240}\pi^4\right)\;.
\end{equation}
The phase-space parametrization in  App.~\ref{sec:ps_factorization_rad_fi_nnlo},
in combination with the sector decomposition of Refs.~\cite{Czakon:2010td,Czakon:2011ve}
can be used to perform these integrals numerically. These results are presented
in Sec.~\ref{sec:numerics}.

\subsubsection{The NNLO case -- Emission of a \texorpdfstring{$q\bar{q}$}{qq} pair}
\label{sec:overlap_removal_nnlo_qq}
The emission of a quark-antiquark pair from a pair of scalar radiators
is described by the insertion operator~\cite{Catani:1999ss,Campbell:2025lrs}
\begin{equation}\label{eq:scalar_emission_quark_pair_2_cs}
  \begin{split}
  \mathcal{S}_{q\bar{q}}(\{p\};q_1,q_2) = &\; \sum_{i,k}
    \hat{\bf T}_i\hat{\bf T}_k\,T_R\,\mathcal{S}^{(q\bar{q})}_{i;k}(q_1,q_2)\;.
  \end{split}
\end{equation}
As in the case of gluon emissions discussed above, it is possible to
eliminate gauge-dependent terms with the help of color conservation.
This leads to a dipole radiator that resembles Eq.~(97) of
Ref.~\cite{Catani:1999ss} in the double-soft limit and exposes
the collinear structure of the $g\to q\bar{q}$ splitting
as well as the gluon spin correlations:
\begin{equation}\label{eq:scalar_emission_quark_pair}
  \begin{split}
  \mathcal{S}_{i;k}^{(q\bar{q})}(q_1,q_2) = &\;
    -\frac{2}{s_{i12}\, s_{k12}}
    \bigg(\frac{2s_{ik}}{s_{12}}-1\bigg)
    +\bigg(\frac{s_{i2}-s_{i1}}{s_{i12}s_{12}}
      -\frac{s_{k2}-s_{k1}}{s_{k12}s_{12}}\bigg)^2\;.
  \end{split}
\end{equation}
In this form, we recognize the second term as a product of the scalar
production and decay current of the intermediate gluon with momentum
$p_{12}^\mu$. The singularities are made explicit by using the notation
of Eq.~\eqref{eq:azimuthal_correlation_polvec}
\begin{equation}\label{eq:scalar_emission_quark_pair_2}
  \begin{split}
    \mathcal{S}^{(q\bar{q})}_{i;k}(q_1,q_2)=&\;
    -\frac{s_{ik}}{s_{i12}\, s_{k12}}\frac{4}{s_{12}}
    \bigg(1-4\big(\epsilon_{ik,\widetilde{12}}\,\epsilon_{12,12}\big)^2\bigg)
    +\frac{2}{s_{i12}s_{k12}}\;,
  \end{split}
\end{equation}
where $\tilde{q}_{12}^\mu=q_{12}^\mu+s_{12}/s_{ik}\,p_{ik}^\mu$.
The mapping to the individual collinear sectors can be performed
by an angular partial fractioning of the form of
Eq.~\eqref{eq:one-gluon_radiator_part}, leading to partitioned
radiators $\mathcal{S}^{(q\bar{q}),(i)}_{i;k}$ which are defined
as in Eq.~\eqref{eq:two-gluon_radiator_nab_part_1}.
The phase-space integrals of the complete subtraction term, in the
notation of Eq.~\eqref{eq:def_integrals}, are given by
(assuming $i\neq k$)
\begin{eqnarray}\label{eq:integral_Sqqik}
  {\cal I}_{i;k}^{(q\bar q)}&=&
  \frac{1}{6\epsilon^3}+\frac{17}{18\epsilon^2}+\frac{1}{\epsilon} \left(\frac{1063}{216}-\frac{11}{36} \pi^2\right)
   +\left(\frac{31009}{1296}-\frac{67}{9} \zeta_3-\frac{169}{108} \pi^2\right)\;,
\\\label{eq:integral_Sqqii}
  {\cal I}_{i;i}^{(q\bar q)}&=&
  -\frac{1}{8\epsilon} -\frac{17}{16} \;.
\end{eqnarray}

\subsection{NLO-like subtraction terms}
\label{sec:nlo_partial_fractioning}
The angular partitioning in Eq.~\eqref{eq:one-gluon_radiator_part}
is insufficient for the subtraction terms of type $\big[\mathcal{M}_{n+1}
  -\mathcal{M}_n\otimes S_1\big]\otimes S_1$ in Eq.~\eqref{eq:structure_2}.
Before a partial fractioning can be applied in this case, one needs to
identify a hierarchy of the emissions. We use the following
decomposition, which is inspired by a parton shower implementation
of double-soft higher-order corrections~\cite{Dulat:2018vuy}\footnote{
  Equation~\eqref{eq:two-gluon_radiator_part_nlo_1} is not used for the
  abelian radiators in Eqs.~\eqref{eq:simplified_abelian_radiators}.}:
\begin{equation}\label{eq:two-gluon_radiator_part_nlo_1}
    (S_1\otimes S_1)_{i,k;l,m}=\big(S_1\otimes S_1)_{i,k;l,m}^{(1)}+
    (\{i,l,1\}\leftrightarrow\{k,m,2\})\;,
\end{equation}
where
\begin{equation}\label{eq:two-gluon_radiator_part_nlo}
    \big(S_1\otimes S_1)_{i,k;l,m}^{(1)}=
    \frac{s_{k1}s_{m1}}{s_{k2}s_{m2}+s_{i1}s_{l1}}\,
    \big(S_1\otimes S_1)_{i,k;l,m}\;,
\end{equation}
and where the indices $i$, $k$, $l$ and $m$ are suitably chosen labels
that match the type of the individual counterterms (e.g.\ $i=k$ in the
case of a pure double-collinear splitting remainder).
For scalar radiators, Eq.~\eqref{eq:two-gluon_radiator_part_nlo} can be
interpreted as an ordering of the gluons according to their transverse momenta,
measured relative to the emitting dipoles in the dipole center-of-mass frames.

\subsection{Spin correlations}
\label{sec:spin_correlations}
Gluon polarization presents certain complications for NLO and NNLO
subtraction. In Refs.~\cite{Campbell:2025lrs,Hoche:2025anb}
it was shown that a straightforward understanding of its effects
can be achieved by considering the quanta of the gluon field as being
emitted and absorbed by QCD color dipoles. In this section, we use
these ideas to define an improved method for the construction of
polarization tensors, suitable for NLO and NNLO subtraction.

\subsubsection{Polarization vectors for color monopoles}
It is well known that axial gauges simplify the computation of
collinear splitting functions and related quantities~\cite{
  Frenkel:1976bia,Dokshitzer:1978hw,Amati:1978wx,Amati:1978by,
  Ellis:1978sf,Ellis:1978ty,Kalinowski:1980ju,Kalinowski:1980wea,
  Humpert:1980te}. The polarization tensor of an axial gauge
\begin{equation}\label{eq:axial_gauge}
    d^{\mu\nu}(q,n)=-g^{\mu\nu}+\frac{q^\mu n^\nu+q^\nu n^\mu}{nq}
    -\frac{n^2\,q^\mu q^\nu}{(nq)^2}\;,
\end{equation}
satisfies the physical requirements for on-shell gluons, namely
$d^\mu_{\;\;\mu}(p,n)=D-2$ and $p_\mu d^{\mu\nu}(p,n)=0$.
A computational technique that bears resemblance to the use of an
axial gauge, but has a different origin, was introduced in the context
of QED~\cite{Fried:1958zz,Grammer:1973db} and has also been used in the
analysis of IR divergences of QCD amplitudes~\cite{Libby:1978qf,Ellis:1978ty}.
Both mimic the effects of coherent gluon emission off particles that are
omitted from the explicit computation, but whose presence is required
by color charge conservation. This is better understood when factorizing
Eq.~\eqref{eq:axial_gauge}:
\begin{equation}\label{eq:classical_axial_gauge}
  d^{\mu\nu}(q,n)=-D^\mu_{\;\;\rho}(q,n)D^{\nu\rho}(q,n)\;,
  \qquad\text{where}\qquad
  D^{\mu\nu}(q,n)=-g^{\mu\nu}+q^\mu\,\frac{n^\nu}{nq}\;.
\end{equation}
It is now apparent how the axial gauge acts on individual diagrams:
When the vector boson momentum, $q^\mu$, in the second term of $D^{\mu\nu}$
is contracted with a vertex factor $(2p+q)^\mu$ or $\gamma^\mu$,
it triggers the elementary Ward identity and pinches the propagator,
decoupling the vector boson described by the polarization sum
from the original charged particle. The same boson is instead coupled
to a Wilson line of velocity $n^\mu$, and the corresponding propagator
denominator is added. Charge conservation guarantees -- in the sum over
all such attachments -- the correct charge of the Wilson line.
In this manner, one is able to construct the dipole needed for
the proper definition of a gluon polarization tensor, even in the case
of only a single well-defined velocity. This corresponds in particular
to the situation encountered in the collinear limit, where the splitting
parton is incorrectly viewed as a QCD monopole radiating other partons.

\subsubsection{Spin-correlated subtraction terms}
Consider now an NLO counterterm comprised of an $e^+e^-\to q\bar{q}g$
underlying Born process, and a $g\to q\bar{q}$ or a $g\to gg$ splitting
function. In both cases, we will have a component of the splitting function
proportional to the dyadic product of two scalar interaction terms of the
form $d^\mu_{\;\nu}(q_{12},n)(q_1-q_2)^\nu$, which arise from the ``decay''
of the gluon~\cite{Bern:1993tz,Campbell:2025lrs,Hoche:2025anb}.
The polarization tensor leading to these expressions should in fact be factorized,
with one of the terms in Eq.~\eqref{eq:classical_axial_gauge} attributed to the
production of the gluon, and one to its splitting. Schematically,
\begin{equation}
  \bigg|\Big[ \mathcal{A}_{q\bar{q}g}^\mu(\tilde{q}_{12},\ldots)
  D_\mu^{\;\,\rho}(q_{12},n)\Big]\frac{1}{q_{12}^2}
  \Big[D^\nu_{\ \rho}(q_{12},n)\,(q_1-q_2)_{\nu}\Big]\bigg|^2\;,
\end{equation}
where $q_1^\mu$ and $q_2^\mu$ are the momenta of the gluon decay products, and
$q_{12}^\mu=q_1^\mu+q_2^\mu$. The quantity $\mathcal{A}_{q\bar{q}g}^\mu(\tilde{q}_{12})$
is the underlying Born amplitude. It is related to the terms in
Eq.~\eqref{eq:structure} as $\mathcal{M}_3=\mathcal{A}_3^\mu(\tilde{q}_{12})
\mathcal{A}_3^{\nu*}(\tilde{q}_{12})\,d_{\mu\nu}(\tilde{q}_{12},n)$.
Crucially, it depends on the \textit{on-shell} gluon momentum, $\tilde{q}_{12}^\mu$,
that is obtained from $q_{12}^\mu$ and the remaining momenta by the momentum mapping
algorithm of the infrared subtraction, cf.\ Sec.~\ref{sec:preliminaries}.
Therefore, if one uses the tensor $D_\mu^{\;\,\rho}(q_{12},n)$, 
one assigns an incorrect external gluon momentum to the polarization
sum associated with the underlying Born matrix element.

Naively, this approach seems to be a viable solution, because 
$D_\mu^{\;\,\rho}(q_{12},n)$ tends to $D_\mu^{\;\,\rho}(\tilde{q}_{12},n)$
in the limit of zero gluon virtuality. However, the on-shell subtractions
$\mathcal{M}_2\otimes S_1$ in the third term on the right-hand side
of Eq.~\eqref{eq:structure_2} are required to cancel $\mathcal{M}_3$
when the gluon with momentum $\tilde{q}_{12}^\mu$ is unresolved in the
three-parton configuration. This situation can arise even if $q_{12}^2$
remains finite. Working with $D_\mu^{\;\,\rho}(q_{12},n)$ in this case
leads to a structurally different gauge-dependence in the terms containing
$\mathcal{M}_2\otimes S_1$ compared to the ones containing $\mathcal{M}_3$.
Once $\tilde{q}_{12}^\mu$ becomes unresolved, this mismatch generates a
gauge-dependent pole. In our method, the problem is solved by consistently
using the partial polarization tensor $D_\mu^{\;\,\rho}(\tilde{q}_{12},n)$
of the three-parton configuration.
For terms involving $\mathcal{M}_3$, we use
\begin{equation}\label{eq:K_decomposition}
  \bigg|\Big[ \mathcal{A}_{q\bar{q}g}^\mu(\tilde{q}_{12},\ldots)
  D_\mu^{\;\,\rho}(\tilde{q}_{12},n)\Big]\frac{1}{q_{12}^2}
  \Big[D^\nu_{\ \rho}(q_{12},n)\,(q_1-q_2)_{\nu}\Big]\bigg|^2\;,
\end{equation}
with the same technique applied to the counterterms of type
$\mathcal{M}_2\otimes S_1$.

Consistency of the infrared subtraction in the double-collinear
limits requires that the momentum mapping be identical between the
on-shell counterterms of the form $\mathcal{M}_2\otimes S_1\otimes S_1$,
and the off-shell terms of the form $\mathcal{M}_2\otimes S_2$
in Eq.~\eqref{eq:structure}. In order to achieve a local cancellation
of polarization dependent singularities, we evaluate $\mathcal{M}_2\otimes S_2$
in the same axial gauge as $\mathcal{M}_2\otimes S_1\otimes S_1$.
We note that this is consistent with the parton-shower approach
presented in Ref.~\cite{Hoche:2025anb}, which only considered the
on-shell scenario and is therefore independent of the momentum mapping.

\section{Numerical results}
\label{sec:numerics}
We validate the subtraction by analyzing the remainders of the double-real
corrections to $e^+e^-\to jj$ in the infrared limits. The double soft and triple
collinear regions constitute the genuinely novel singular configurations encountered
at NNLO. To obtain numerically stable results, we employ the \textsc{Pepper} event
generator~\cite{Bothmann:2021nch,Bothmann:2023siu,Bothmann:2023gew}, including
the techniques described in Ref.~\cite{Bothmann:2024hgq}. In addition, we use
arbitrary-precision arithmetic as implemented by
\texttt{mp++}~\cite{francesco_biscani_2024_14357329}.

\begin{figure}[t]
    \centering
    \includegraphics[width=.49\textwidth]{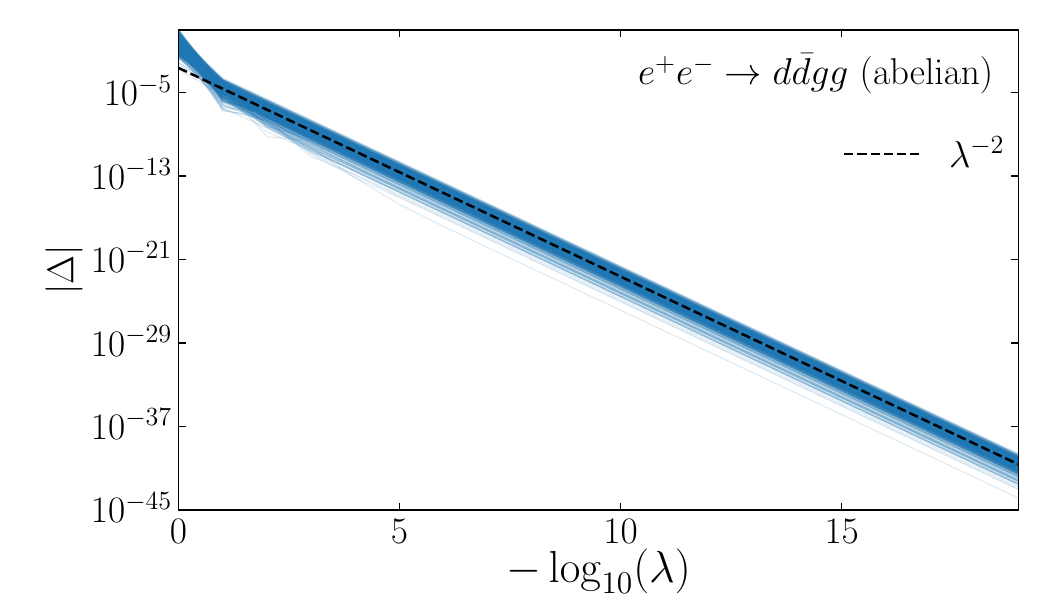}  
    \includegraphics[width=.49\textwidth]{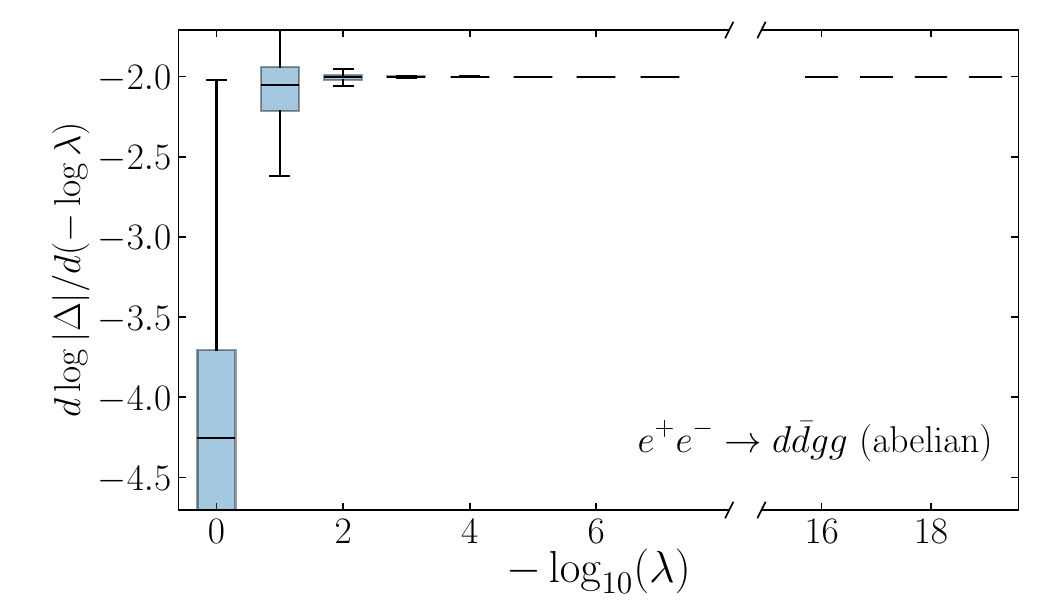}
    \caption{Left: Scaling behavior of the subtracted $e^+e^-\to q\bar{q}gg$
    squared matrix element in the double-soft limit. Each line corresponds to a
    randomly generated phase-space configuration, which is scaled into the limit
    by means of the algorithms in Refs.~\cite{Asteriadis:2020gzh,Bothmann:2024hgq}. 
    Right: Numerically computed slope of $\log|\Delta|$ versus $\log(1/\lambda)$ as a function
    of the scaling parameter $\lambda$ for the double-soft limit of the abelian component 
    of $e^+e^-\to q\bar{q}gg$.} 
    \label{fig:limits}
\end{figure}
Figure~\ref{fig:limits} shows the relative remainder,
\begin{equation}
    \Delta = \frac{|\mathcal{M}_{RR}|}{\mathcal{M}_4}\;,
\end{equation}
as a function of the logarithm of an infrared scaling parameter, $\lambda$.
Each line in the plot on the left corresponds to one of 250 randomly
generated double-real phase-space points, which is scaled into the double-soft
limit using the methods described in Refs.~\cite{Asteriadis:2020gzh,Bothmann:2024hgq}.
As $\lambda \to 0$, the matrix element diverges as $\lambda^{-2}$
in the single-unresolved regions and as $\lambda^{-4}$ in the double
unresolved regions. A subtraction that correctly captures the leading
singular behavior will cause $\Delta$ to vanish proportional to $\lambda$,
making the slopes of the individual trajectories a clean and unambiguous
diagnostic of the correctness of the method. These slopes approach $-2$
for the abelian component of $\Delta(e^+e^-\to q\bar{q}gg)$ shown in
Fig.~\ref{fig:limits}. Our method thus not only captures the leading,
but also the sub-leading behavior of the full matrix element in the
double-soft limit. The right panel shows the distribution of the
derivatives as a function of $\log(1/\lambda)$ for $10^4$ phase-space
points. It highlights that, after a few orders of magnitude in $\lambda$,
all slopes stabilize at $-2$, with a near negligible variation among
the individual starting configurations.

Table~\ref{tab:limits} shows the average derivative and its uncertainty
for all structurally different limits, and all types of partonic
sub-processes. We separate the partonic channels including two quarks and
two gluons into their abelian and non-abelian color structures.
We find that the derivatives not shown in Fig.~\ref{fig:limits}
stabilize at a value of $-1$, indicating that our infrared subtraction
correctly captures the leading power in these cases. The small
statistical uncertainties associated with the results again highlight
the stability of the scaling when approaching each limit. These
variations are determined from a sample of $10^4$ phase space points.

\begin{table}[t]
    \centering
        \begin{tabular}{lcccc}
        \toprule
        Limit & $q\bar{q}Q\bar{Q}$ & $q\bar{q}q\bar{q}$ & $q\bar{q}gg$ (ab) & $q\bar{q}gg$ (nab) \\
        \midrule
        Single Soft         & --                & --                & $-1 \pm 10^{-17}$ & $-1 \pm 10^{-18}$ \\
        Double Collinear    & $-1 \pm 10^{-17}$ & $-1 \pm 10^{-18}$ & $-1 \pm 10^{-17}$ & $-1 \pm 10^{-17}$ \\
        Double Soft         & $-1 \pm 10^{-18}$ & $-1 \pm 10^{-19}$ & $-2 \pm 10^{-18}$ & $-1 \pm 10^{-17}$ \\
        Triple Collinear    & $-1 \pm 10^{-19}$ & $-1 \pm 10^{-19}$ & $-1 \pm 10^{-19}$ & $-1 \pm 10^{-19}$ \\
        Soft Collinear      & --                & --                & $-2 \pm 10^{-18}$ & -- \\
        Double Collinear x2 & --                & --                & $-2 \pm 10^{-19}$ & -- \\
        \bottomrule
        \end{tabular}
    \caption{Numerically computed slope of $\log|\Delta|$ at $-\log_{10}(\lambda)=20$ for all
    types of single- and double-unresolved limits and processes considered in this work.
    Dashes indicate limits in which the squared matrix element has no leading pole.
    } 
    \label{tab:limits}
\end{table}
To test the numerical performance that can be expected of our method
we compute the total double-real contribution to the cross section
for $e^+e^-\to q\bar{q}$ at the $Z$ pole. For simplicity, we do not use an
electron structure function. Since the precise numerical value of the
subtracted double-real corrections depends on the subtraction scheme,
we only quote the uncertainty of the result. With $5.1\cdot10^{8}$
phase-space points, we obtain a $1\sigma$-uncertainty of $654$fb,
which is less than $2\cdot 10^{-5}$ relative to the Born cross section.
Achieving the expected precision of a Tera-$Z$ option of a potential
FCC-ee is thus possible with a relatively modest amount of computing.

\begin{figure}[t]
  \centering
  \includegraphics[width=.495\textwidth]{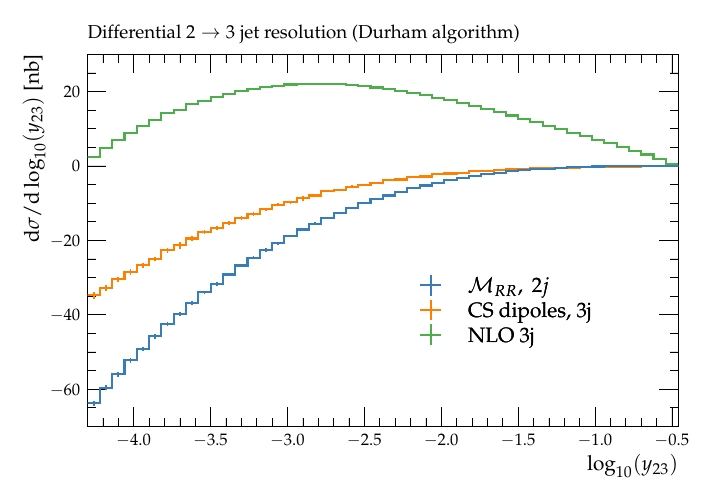}
  \includegraphics[width=.495\textwidth]{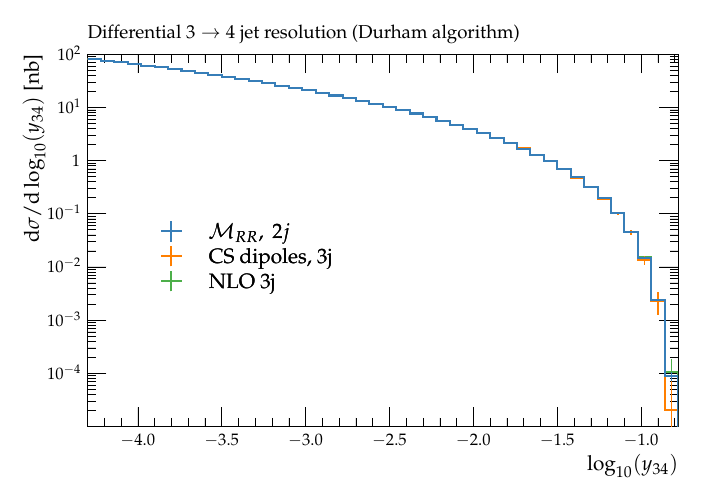}
  \caption{Differential jet rates in the Durham algorithm at $\sqrt{s}=91.2$~GeV.
    Left: $2\to 3$-jet rate. Right: $3\to 4$-jet rate. Results for $\mathcal{M}_{RR}$
    in Eq.~\eqref{eq:structure} are compared to NLO subtraction results (labeled
    `CS dipoles, 3j') for $e^+e^-\to 3$~jets with $y_{23}>3\cdot 10^{-5}$, and using
    the Catani-Seymour subtraction scheme~\cite{Catani:1996vz}. We also include
    the corresponding complete NLO 3-jet prediction.
    \label{fig:jet_rates}}
\end{figure}

Figure~\ref{fig:jet_rates} shows two distributions of interest
in $e^+e^-\to q\bar{q}$, the differential $2\to 3$ and $3\to 4$ jet rates in
the Durham algorithm~\cite{Catani:1991hj}. In a NNLO fixed-order calculation,
$y_{34}$ is predicted at LO accuracy, while $y_{23}$ is predicted at NLO
accuracy. The two distributions therefore only probe the performance
of our calculation in simple (NLO-like) situations. However, together
with the information on the statistical precision of the total cross section,
we can infer that the numerical convergence is sufficient to enable precision
physics at a potential future electron-positron collider.
Finally, we verify that the integration of the scalar counterterms
can be carried out numerically. This involves the parametrization of the
two-emission phase space from  App.~\ref{sec:ps_factorization_rad_fi_nnlo},
combined with the sector decomposition introduced in
Refs.~\cite{Czakon:2010td,Czakon:2011ve}. Note that we only use the
mapping of the sector variables from these references, not the original
kinematics reconstruction. This is essential to factorize the Born
phase-space exactly. For the numerical evaluation we make use of
pySecDec~\cite{Borowka:2017idc,Borowka:2018goh,Heinrich:2021dbf,
  Heinrich:2023til}, and we use the selector function from
Ref.~\cite{Czakon:2011ve} with $\alpha=0$. We have cross-checked these results
using the standalone code available at \url{https://gitlab.com/shoeche/cocos}.
In back-to-back kinematics, the $q\bar{q}$ radiators give the results
in Tab.~\ref{tab:secdec_quarkantiquark}, the abelian two-gluon radiators
give the results in Tab.~\ref{tab:secdec_abelian} and the non-abelian
two-gluon radiators give the results in Tab.~\ref{tab:secdec_non_abelian}.
Each column of the tables corresponds to one of the sectors. The final
column shows their sum, relative to the analytic result. For all terms
implemented using partial fractioning, we present two versions of the
integrals, once the complete result and once including the partial
fractioning introduced in Sec.~\ref{sec:overlap_removal}. This confirms
that the integrals have the expected divergence structure. For example,
${\cal I}_{i,i;k,k}^{({\rm ab}),(i,i)}$ is singular only in the
region described by sectors $S^{I}_1$-$S^{I}_5$, while
${\cal I}_{i,i;k,k}^{({\rm ab}),(i,k)}$ is singular only in the
region described by sectors $S^{II}_1$ and $S^{II}_2$
(see Tab.~\ref{tab:mapping_sector_variables} in
  App.~\ref{sec:ps_factorization_rad_fi_nnlo} for a definition
  of the individual sectors).
Note that, in the back-to-back case, the symmetry fixes the angular partition
weights to integrate to one half, while in the abelian two-gluon radiator,
where we partition with a product of two angular weights, the two integrals
instead sum to one half. This can be observed clearly in
Tabs.~\ref{tab:secdec_quarkantiquark}-\ref{tab:secdec_non_abelian}.
\begin{table}[t]
    \centering
    \begin{tabular}{ll@{\hskip 2mm}cccccccc}
        \toprule
        Integral 
        & $\mathcal{O}$
        & $S^I_1$ & $S^I_2$ & $S^I_3$ & $S^I_4$ & $S^I_5$ & $S^{II}_1$ & $S^{II}_2$ & Ratio \\
        \midrule
        \multirow{2}{*}{${\cal I}_{i,i}^{(q\bar q)}$}
        & $\epsilon^{-1}$
        & -0.0216834(5) & -0.0223860(5) & -0.0364991(7) & -0.019984(1) & -0.024448(1) & 0 & 0 & 1.00000(2) \\
        & $\epsilon^{0}$
        & -0.196006(4) & -0.248492(4) & -0.400420(7) & -0.147282(8) & -0.18715(1) & 0.0155611(3) & 0.101289(3) & 1.00000(2) \\
        \midrule
        \multirow{4}{*}{${\cal I}_{i,k}^{(q\bar q)}$}
        & $\epsilon^{-3}$
        & 0 & 0 & 0 & 1/12 & 1/12 & 0 & 0 & 1\\
        & $\epsilon^{-2}$
        & -0.151813(5) & -0.198946(6) & -0.058384(2) & 0.69788(2) & 0.65569(2) & 0 & 0 & 0.99998(3) \\
        & $\epsilon^{-1}$
        & -1.63192(3) & -2.76223(4) & -0.66430(1) & 3.3448(2) & 2.6891(1) & 0.56399(2) & 0.36586(2) & 0.9998(1) \\
        & $\epsilon^{0}$
        & -9.2643(2) & -21.3178(3) & -4.12296(7) & 13.4332(9) & 7.9172(9) & 8.5579(4) & 4.3281(3) & 1.006(3) \\
        \midrule
        \multirow{4}{*}{${\cal I}_{i,k}^{(q\bar q),(i)}$}
        & $\epsilon^{-3}$
        & 0 & 0 & 0 & 1/24 & 1/24 & 0 & 0 & 1/2 \\
        & $\epsilon^{-2}$
        & -0.075907(2) & -0.099473(3) & -0.0291923(8) & 0.34894(1) & 0.327843(8) & 0 & 0 & 0.49999(1) \\
        & $\epsilon^{-1}$
        & -0.81596(1) & -1.38111(2) & -0.332149(6) & 1.67239(8) & 1.34455(7) & 0.281997(9) & 0.182928(9) & 0.49992(6) \\
        & $\epsilon^{0}$
        & -4.63215(8) & -10.6589(2) & -2.06148(3) & 6.7166(5) & 3.9586(5) & 4.2790(2) & 2.1641(1) & 0.503(2) \\
        \bottomrule
    \end{tabular}
    \caption{Numerical comparison of the integrated, scalar quark-antiquark radiators in the notation 
    of Eq.~\eqref{eq:def_integrals} and using the sector definitions of Ref.~\cite{Czakon:2010td}
    (see also Tab.~\ref{tab:mapping_sector_variables}).
    The last column shows the ratio of the sum of all sectors to the analytic results in Eqs.~\eqref{eq:integral_Sqqik} and~\eqref{eq:integral_Sqqii}.
    For the integrals from partial fractioned integrands, we still use ${\cal I}_{i;k}^{(q\bar q)}$
    as a reference.}
    \label{tab:secdec_quarkantiquark}
\end{table}

\begin{table}[t]
    \centering
    \begin{tabular}{ll@{\hskip 2mm}cccccccc}
        \toprule
        Integral 
        & $\mathcal{O}$
        & $S^I_1$ & $S^I_2$ & $S^I_3$ & $S^I_4$ & $S^I_5$ & $S^{II}_1$ & $S^{II}_2$ & Ratio\\
        \midrule
        \multirow{5}{*}{${\cal I}_{i,i;k,k}^{\rm(ab)}$}
        & $\epsilon^{-4}$
        & 1/16 & 1/48 & 1/24 & 0 & 0 & 1/24 & 1/12 & 1\\
        & $\epsilon^{-3}$
        & 0.293324(8) & 0.097771(5) & 0.195555(6) & -0.043317(4) & -0.043327(4) & 0.16668(1) & 0.33335(1) & 1.00004(2) \\
        & $\epsilon^{-2}$
        & 0.41263(4) & 0.19826(4) & 0.21430(4) & -0.13224(3) & -0.13224(3) & -0.09662(6) & 0.33484(4) & 1.0000(1) \\
        & $\epsilon^{-1}$
        & -0.4470(2) & 0.5827(2) & -1.0301(2) & 0.2993(1) & 0.2995(2) & -2.2912(4) & -0.6309(2) & 1.0002(2) \\
        & $\epsilon^{0}$
        & -2.4961(7) & 4.072(2) & -6.568(2) & 2.7304(5) & 2.7315(6) & -7.701(3) & -2.299(2) & 1.0002(5) \\
        \midrule
        \multirow{5}{*}{${\cal I}_{i,i;k,k}^{({\rm ab}),(i,i)}$}
        & $\epsilon^{-4}$
        & 1/32 & 1/96 & 1/48 & 0 & 0 & 0 & 0 & 1/4 \\
        & $\epsilon^{-3}$
        & 0.189988(8) & 0.063325(6) & 0.126652(8) & -0.021662(2) & -0.021660(2) & -0.057759(3) & -0.028881(1) & 0.25000(1) \\
        & $\epsilon^{-2}$
        & 0.54079(3) & 0.19940(4) & 0.34124(3) & -0.11582(1) & -0.11581(1) & -0.53250(2) & -0.036974(9) & 0.35090(8) \\
        & $\epsilon^{-1}$
        & 1.0498(2) & 0.6078(2) & 0.4415(2) & -0.24829(4) & -0.24830(4) & -2.6004(1) & 0.94579(6) & 0.0162(1) \\
        & $\epsilon^{0}$
        & 2.046(1) & 2.675(3) & -0.626(1) & -0.2313(2) & -0.2312(2) & -9.7481(8) & 7.2490(5) & -0.1189(4) \\
        \midrule
        \multirow{5}{*}{${\cal I}_{i,i;k,k}^{({\rm ab}),(i,k)}$}
        & $\epsilon^{-4}$
        & 0 & 0 & 0 & 0 & 0 & 1/48 & 1/24 & 1/4 \\
        & $\epsilon^{-3}$
        & -0.043321(2) & -0.0144414(8) & -0.028880(1) & 0 & 0 & 0.14111(1) & 0.19555(1) & 0.25001(1) \\
        & $\epsilon^{-2}$
        & -0.33444(1) & -0.100326(7) & -0.234106(9) & 0.049691(3) & 0.049693(3) & 0.48426(4) & 0.20435(3) & 0.14910(7) \\
        & $\epsilon^{-1}$
        & -1.27342(4) & -0.31662(5) & -0.95666(3) & 0.39796(2) & 0.39798(2) & 1.4553(2) & -1.2614(2) & 0.4839(1) \\
        & $\epsilon^{0}$
        & -3.2951(3) & -0.6362(9) & -2.6592(3) & 1.59668(6) & 1.59675(5) & 5.901(2) & -8.399(1) & 0.6187(3) \\
        \midrule
        \multirow{2}{*}{${\cal I}_{i,i;k}^{\rm(ab)}$}
        & $\epsilon^{-1}$
        & -0.01896(1) & -0.02015(1) & -0.03242(2) & -0.013287(6) & -0.016212(6) & 0 & 0 & 1.0000(2) \\
        & $\epsilon^{0}$
        & -0.1561(1) & -0.2315(2) & -0.3619(3) & -0.07571(5) & -0.10252(5) & 0.02207(3) & 0.1446(1) & 1.0003(5) \\
        \midrule
        \multirow{2}{*}{${\cal I}_{i;i}^{\rm(ab)}$}
        & $\epsilon^{-1}$
        & -0.0216834(5) & -0.0223860(5) & -0.0364990(8) & -0.019984(1) & -0.024447(1) & 0 & 0 & 1.00000(2) \\
        & $\epsilon^{0}$
        & -0.174323(4) & -0.226106(4) & -0.363920(6) & -0.127298(7) & -0.162701(9) & 0.0155611(3) & 0.101289(3) & 1.00000(2) \\
        \midrule
        ${\cal I}_{i;k}^{\rm(ab)}$ 
        & $\epsilon^{0}$
        & 0.004563(1) & 0.0019061(7) & 0.004211(1) & 0.015988(3) & 0.017010(3) & 0.0001819(3) & 0.008023(3) & 0.9999(1) \\
        \bottomrule
    \end{tabular}
    \caption{Numerical comparison of the integrated, scalar abelian two-gluon radiators in the notation 
    of Eq.~\eqref{eq:def_integrals} and using the sector definitions of Ref.~\cite{Czakon:2010td}
    (see also Tab.~\ref{tab:mapping_sector_variables}).
    The last column shows the ratio of the sum of all sectors to the analytic results in Eqs.~\eqref{eq:integral_Sab_iikk}-\eqref{eq:integral_Sab_ik}.
    For the integrals from partial fractioned integrands, we still use ${\cal I}_{i,i;k,k}^{\rm(ab)}$
    as a reference.}\label{tab:secdec_abelian}
\end{table}

\begin{table}[t]
    \centering
    \begin{tabular}{ll@{\hskip 2mm}cccccccc}
        \toprule
        Integral 
        & $\mathcal{O}$
        & $S^I_1$ & $S^I_2$ & $S^I_3$ & $S^I_4$ & $S^I_5$ & $S^{II}_1$ & $S^{II}_2$ & Ratio\\
        \midrule
        \multirow{5}{*}{${\cal I}_{i;k}^{({\rm nab},s.o.)}$}
        & $\epsilon^{-4}$
        & 0 & 0 & 0 & 1/4 & 1/4 & 0 & 0 & 1\\
        & $\epsilon^{-3}$
        & -0.34658(2) & -0.34658(2) & 0 & 1.34659(4) & 1.34659(4) & 0 & 0 & 1.00001(3) \\
        & $\epsilon^{-2}$
        & -2.8231(1) & -3.4054(2) & 0.58227(6) & 2.2995(2) & 2.2997(2) & 1.64498(9) & 0 & 1.0004(6) \\
        & $\epsilon^{-1}$
        & -11.1988(5) & -17.751(1) & 6.5513(6) & -4.6191(6) & -4.6177(6) & 19.502(1) & -3.9068(3) & 1.0000(1) \\
        & $\epsilon^{0}$
        & -28.379(2) & -68.731(6) & 40.343(5) & -41.844(2) & -41.840(2) & 121.691(9) & -45.065(6) & 1.0001(2) \\
        \midrule
        \multirow{5}{*}{${\cal I}_{i;k}^{\rm(nab)}$}
        & $\epsilon^{-4}$
        & 0 & 0 & 0 & 1/4 & 1/4 & 0 & 0 & 1 \\
        & $\epsilon^{-3}$
        & -0.34658(2) & -0.34658(2) & 0 & 1.80491(4) & 1.80489(4) & 0 & 0 & 0.99999(2) \\
        & $\epsilon^{-2}$
        & -3.4649(1) & -4.3255(2) & 0.18702(4) & 7.0902(2) & 6.8164(2) & 1.64488(9) & 0 & 1.00001(5) \\
        & $\epsilon^{-1}$
        & -18.9954(6) & -30.890(1) & 1.5569(4) & 22.437(1) & 18.386(1) & 21.816(1) & -1.5611(2) & 0.9999(2) \\
        & $\epsilon^{0}$
        & -79.238(2) & -171.659(8) & 5.948(3) & 71.786(5) & 38.821(5) & 158.64(1) & -14.207(3) & 0.995(2) \\
        \midrule
        \multirow{5}{*}{${\cal I}_{i;k}^{({\rm nab}),(i)}$}
        & $\epsilon^{-4}$
        & 0 & 0 & 0 & 1/8 & 1/8 & 0 & 0 & 1/2 \\
        & $\epsilon^{-3}$
        & -0.173292(8) & -0.17329(1) & 0 & 0.90246(2) & 0.90245(2) & 0 & 0 & 0.50000(1) \\
        & $\epsilon^{-2}$
        & -1.73243(6) & -2.1628(1) & 0.09351(2) & 3.54509(9) & 3.4082(1) & 0.82250(4) & 0 & 0.50001(2) \\
        & $\epsilon^{-1}$
        & -9.4977(3) & -15.4448(6) & 0.7784(2) & 11.2184(5) & 9.1928(5) & 10.9099(5) & -0.7807(1) & 0.50005(9) \\
        & $\epsilon^{0}$
        & -39.619(1) & -85.830(4) & 2.974(1) & 35.893(2) & 19.411(2) & 79.347(4) & -7.105(2) & 0.4999(7) \\
        \midrule
        \multirow{3}{*}{${\cal I}_{i;i}^{\rm(nab)}$}
        & $\epsilon^{-2}$
        & 0 & 0 & 0 & 1/2 & 1/2 & 0 & 0 & 1 \\
        & $\epsilon^{-1}$
        & -0.69316(1) & -0.85419(1) & -0.53209(1) & 4.6929(1) & 4.3864(1) & 0 & 0 & 0.99999(1) \\
          & $\epsilon^{0}$
        & -7.92686(7) & -12.01070(9) & -6.60800(7) & 24.1601(9) & 19.6443(8) & 1.84454(8) & 3.0898(3) & 0.9999(1) \\
        \bottomrule
    \end{tabular}
    \caption{Numerical comparison of the integrated, scalar non-abelian two-gluon radiators in the notation 
    of Eq.~\eqref{eq:def_integrals} and using the sector definitions of Ref.~\cite{Czakon:2010td}
    (see also Tab.~\ref{tab:mapping_sector_variables}).
    The last column shows the ratio of the sum of all sectors to the analytic results in
    Eqs.~\eqref{eq:integral_Snab_ik}-\eqref{eq:interal_Snabso_ik}.
    For the integrals from partial fractioned integrands, we still use ${\cal I}_{i;k}^{\rm(nab)}$
    as a reference.}
    \label{tab:secdec_non_abelian}
\end{table}

\section{Outlook}
\label{sec:outlook}
We have presented an algorithm for double-real radiative corrections
in $e^+e^-\to q\bar{q}$.
The individual subtraction terms are constructed from scalar radiators
and pure splitting functions. The NNLO kinematics mappings are simple
iterations of the NLO mappings.
To complete the calculation, an integration of the infrared counterterms
over the unresolved phase space must be performed. The singularities occurring
in the scalar radiators are the most complicated in this regard, because soft
and collinear regions overlap. We have matched these overlapping terms to
individual double- and triple-collinear regions using the techniques
in Sec.~\ref{sec:overlap_removal}. We have checked that a numerical integration
can be performed in terms of a Laurent series expansion of the integrand, using
the phase-space parametrization discussed in App.~\ref{sec:ps_factorization},
in conjunction with the sectorization techniques in Refs.~\cite{Czakon:2010td,Czakon:2011ve}.
We have also provided analytic results for the back-to-back case.
The pure splitting functions have a simpler pole structure than
the scalar radiators, and we will address their integrals in a future publication.

In order to complete the NNLO calculation for $e^+e^-\to q\bar{q}$,
it also remains to perform a partial fractioning of the one-loop scalar radiator functions
derived in Ref.~\cite{Campbell:2025lrs}, and to derive the corresponding integrated
subtraction counterterms. We will address this problem in a future publication.

\section*{Acknowledgments}
\noindent
We would like to thank Thomas Gehrmann and Raoul R{\"o}ntsch for inspiring
and helpful discussions on fixed-order subtraction algorithms.
We are grateful to Thomas Gehrmann for comments on the manuscript.
This manuscript has been authored by Fermi Forward Discovery Group, LLC
under Contract No. 89243024CSC000002 with the U.S.\ Department of Energy,
Office of Science, Office of High Energy Physics.
This research used resources of the National Energy Research Scientific Computing Center (NERSC), 
a Department of Energy Office of Science User Facility, using NERSC award ERCAP0035011.
The work of J.M.C., S.H.\ and M.K.\ was supported by the U.S. Department of Energy,
Office of Science, Office of Advanced Scientific Computing Research,
Scientific Discovery through Advanced Computing (SciDAC-5) program,
grant “NeuCol”. The configuration of pySecDec was performed
with the help of a Large Language Model~\cite{Anthropic:2026aa}.

\appendix
\section{Kinematics and phase-space factorization}
\label{sec:ps_factorization}
In this appendix we discuss the factorization of the differential $n+2$
particle phase-space element into a differential $n$ particle phase-space
element and the radiative phase space. The starting point is the manifestly
covariant form of the phase-space element in $D=4-2\eps$ space-time dimensions,
\begin{equation}\label{eq:two_body_ps_ddim}
    {\rm d}{\Phi}_2(P;p,q)=\frac{(4\pi^2 N^2)^{\eps}}{16\pi^2}\,
    \frac{((pN)^2-p^2N^2)^{3/2-\eps}}{((pN)(Pp)-p^2(PN))N^2}\,{\rm d}\Omega_{p,N}^{2-2\eps}\;,
\end{equation}
where the vector $N^\mu$ defines a frame of reference in which the solid angle
integral is carried out.

\subsection{NLO kinematics}
\label{sec:ps_factorization_rad_fi}
\begin{figure}[t]
\includegraphics[width=\textwidth]{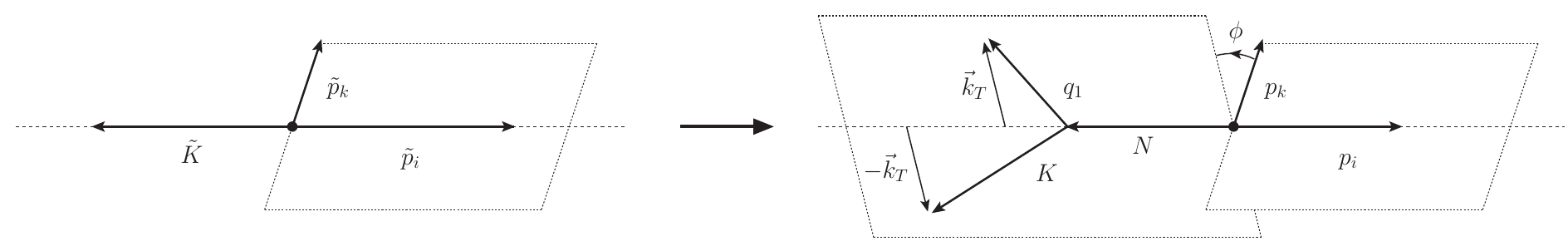}
\caption{Sketch of the kinematics mapping in the identified particle
subtraction algorithm of~\cite{Catani:1996vz}. See the main text for details.
Note that $p_k$ is unaffected by the mapping and only acts as a reference
for the azimuthal angle $\phi$.\label{fig:ff_rad_kinematics}}
\end{figure}
Figure~\ref{fig:ff_rad_kinematics} shows the kinematics mapping we employ
at NLO. At NNLO, the technique is applied twice, with the successive assignment
of the radiator performed according to the algorithms in Sec.~\ref{sec:overlap_removal}.
The basic momentum mapping was introduced in Sec.~5.6 of Ref.~\cite{Catani:1996vz}.
Beginning with Eq.~\eqref{eq:def_n_pi},
\begin{equation}\label{eq:def_n_pi_2}
  p_i^\mu=z\,\tilde{p}_{i}^\mu\;,
  \qquad
  N^\mu=\tilde{K}^\mu+(1-z)\,\tilde{p}_{i}^\mu\;,
\end{equation}
we can define $N^\mu=K^\mu+q_1^\mu$ and use a Sudakov decomposition~\cite{Sudakov:1954sw}
along $\tilde{p}_i^\mu$ and the direction of $N^\mu$ to construct the new momenta.
Using the light-like vector
\begin{equation}
  \bar{N}^\mu=N^\mu-\frac{N^2}{2\tilde{p}_iN}\,\tilde{p}_i^\mu
  =\tilde{K}^\mu-\kappa\,\tilde{p}_i^\mu\;,
  \qquad\text{where}\qquad
  \kappa=\frac{\tilde{K}^2}{2\tilde{p}_i\tilde{K}}\;,
\end{equation}
we can write
\begin{equation}\label{eq:def_pi_K}
    \begin{split}
        q_1^\mu&=v\,\bar{N}^\mu+\frac{1}{v}\frac{{\rm k}_\perp^2}{2\tilde{p}_i\tilde{K}}\,\tilde{p}_i^\mu+k_\perp^\mu\;,
        &\text{where}\qquad
        v&=\frac{p_iq_1}{p_iN}\;,\\
        K^\mu&=(1-v)\,\bar{N}^\mu+\frac{1}{1-v}\frac{{\rm k}_\perp^2+\tilde{K}^2}{\,2\tilde{p}_i\tilde{K}}\,\tilde{p}_i^\mu-k_\perp^\mu\;.
    \end{split}
\end{equation}
Conservation of the $\tilde{p}_i^\mu$ components results in the following
identity for the magnitude of the transverse momentum
\begin{equation}
  \label{eq:def_kt2_identified}
  {\rm k}_\perp^2=2\tilde{p}_i\tilde{K}\,v\,\big[(1-z+\kappa)(1-v)-\kappa\big]\;.
\end{equation}
Substituting this into Eq.~\eqref{eq:def_pi_K} makes
both collinear safety and overall four-momentum conservation manifest.
\begin{equation}\label{eq:pi_K_using_kt2}
    \begin{split}
        q_1^\mu&=(1-z)\,\tilde{p}_i^\mu+v\big(\tilde{K}^\mu-(1-z+2\kappa)\,\tilde{p}_i^\mu\big)+k_\perp^\mu\;,\\
        K^\mu&=\tilde{K}^\mu-v\big(\tilde{K}^\mu-(1-z+2\kappa)\,\tilde{p}_i^\mu\big)-k_\perp^\mu\;.
    \end{split}
\end{equation}
In order to derive the phase-space factorization, we make use
of the relation~\cite{Byckling:1969sx}
\begin{equation}\label{eq:emission_phase_space_fi_rad}
    {\rm d}\Phi_{3}(-K;p_i,q_1,Q)
    ={\rm d}\Phi_{2}(-K;q_1,-N)\,
    \frac{{\rm d}N^2}{2\pi}\,
    {\rm d}\Phi_{2}(-N;p_i,Q)\;,
\end{equation}
where $Q^\mu=\sum_{l\neq i,j}p_l^\mu$ is the sum of all final-state momenta
except the collinear pair $p_i^\mu$ and $q_1^\mu$. Using the techniques of
Ref.~\cite{Assi:2023rbu}, we can relate the phase-space factor
${\rm d}\Phi_2(-N;p_i,Q)$ to the underlying Born phase space.
In the case of massless radiators one obtains
\begin{equation}\label{eq:born_remap_fi_rad}
  \begin{split}
    \frac{{\rm d}\Phi_2(-N;{p}_i,Q)}{
    {\rm d}\Phi_2(-\tilde{K};\tilde{p}_{i},\tilde{Q})}
    =&\;\bigg(\frac{p_i N}{
      \tilde{p}_{i} \tilde{K}}\bigg)^{3-2\eps}\;
    \frac{(\tilde{p}_{i}\tilde{K})^2}{({p}_iN)^2}
    =z^{1-2\eps}\;.
  \end{split}
\end{equation}
The differential two-body phase-space element for the production of
$N^\mu$ and $q_1^\mu$ is given by
\begin{equation}\label{eq:pj_production_ps_ff}
  \begin{split}
    {\rm d}{\Phi}_2(-K;q_1,-N)=&\;\frac{(4\pi^2)^{\eps}}{16\pi^2}\,
    \frac{(q_1N)^{1-2\eps}}{(N^2)^{1-\eps}}\,{\rm d}\Omega_{1,N}^{2-2\eps}
    =\frac{(2\tilde{p}_{i}\tilde{K})^{-\eps}}{(16\pi^2)^{1-\eps}}\,
    \frac{(1-z)^{1-2\eps}}{2(1-z+\kappa)^{1-\eps}}\,{\rm d}\Omega_{1,N}^{2-2\eps}\;,
  \end{split}
\end{equation}
Finally, we combine Eqs.~\eqref{eq:born_remap_fi_rad} and~\eqref{eq:pj_production_ps_ff}
to obtain the single-emission phase space element
\begin{equation}\label{eq:emission_phase_space_if_ml}
  {\rm d}\Phi_{+1}(-\tilde{K},\tilde{p}_{i};q_1)
  =\bigg(\frac{2\tilde{p}_{i}\tilde{K}}{16\pi^2}\bigg)^{1-\eps}\,
  \frac{\big(z(1-z)\big)^{1-2\eps}}{(1-z+\kappa)^{1-\eps}}\,{\rm d}z\,
      \frac{{\rm d}\Omega_{1,N}^{2-2\eps}}{4\pi}\;,
\end{equation}
where we have used the fact that $N^2=2\tilde{p}_i\tilde{K}(1-z+\kappa)$.
This expression agrees with the result derived in Ref.~\cite{Catani:1996vz}.
Using Eq.~\eqref{eq:emission_phase_space_if_ml} to compute the 1-emission
phase-space volume in the back-to-back configuration, $\kappa=-1$, gives
\begin{equation}\label{eq:ps_volume_3}
  \Phi_{+1}(-\tilde{K})=\frac{(4\pi)^{\eps}}{16\pi^2}
  \frac{\Gamma(1-\eps)^2}{\Gamma(3-3\eps)}\,
  \big(\tilde{K}^2\big)^{1-\eps}\;,
\end{equation}
in agreement with App.~A of Ref.~\cite{Gehrmann-DeRidder:2003pne}.

\subsection{NNLO kinematics}
\label{sec:ps_factorization_rad_fi_nnlo}
\begin{figure}[t]
\includegraphics[width=\textwidth]{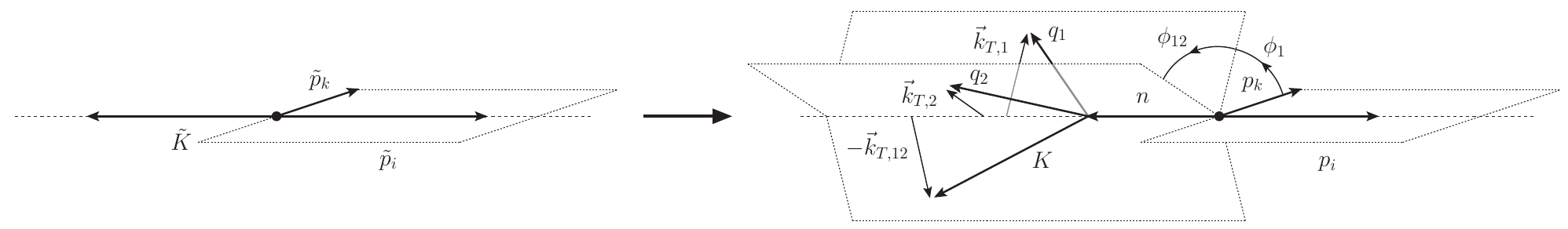}
\caption{Sketch of the two-emission phase space mapping in parametrization~(a).
\label{fig:ff_rad_kinematics_nnlo}}
\end{figure}
The phase-space factorization at NNLO can be achieved in different ways.
In a triple-collinear parametrization, the polar angle of both emitted partons
is typically measured against the direction of $\tilde{p}_i^\mu$, and the
longitudinal recoil is taken by parton $i$. We will begin by deriving
the $D$-dimensional phase-space element in this case.

A sketch of the kinematics is shown in Fig.~\ref{fig:ff_rad_kinematics_nnlo}. 
We employ a kinematical map that is similar to the NLO case in
Sec.~\ref{sec:preliminaries}. Starting with the identified momentum,
$\tilde{p}_i^\mu$, and the recoil momentum, $\tilde{K}^\mu$, we use
Eq.~\eqref{eq:def_n_pi} to define $p_i^\mu$ and $N^\mu$. As we intend
to generate the two new momenta, $q_1^\mu$ and $q_2^\mu$,
we define $N^\mu=K^\mu+q_1^\mu+q_2^\mu$. The Sudakov decomposition
along $\tilde{p}^\mu$ and the direction of $N^\mu$ leads to
\begin{equation}\label{eq:def_pi_K_nnlo}
    \begin{split}
        q_1^\mu&=v_1\,\bar{N}^\mu+\frac{1}{v_1}\frac{{\rm k}_{\perp,1}^2}{
          2\tilde{p}_i\tilde{K}}\,\tilde{p}_i^\mu+k_{\perp,1}^\mu\;,
        &
        v_1&=\frac{p_iq_1}{p_iN}\\
        q_2^\mu&=v_2\,\bar{N}^\mu+\frac{1}{v_2}\frac{{\rm k}_{\perp,2}^2}{
          2\tilde{p}_i\tilde{K}}\,\tilde{p}_i^\mu+k_{\perp,2}^\mu\;,
        &\text{where}\qquad
        v_2&=\frac{p_iq_2}{p_iN}\\
        K^\mu&=(1-v_{12})\,\bar{N}^\mu+\frac{1}{1-v_{12}}\frac{{\rm k}_{\perp,12}^2
        +\tilde{K}^2}{\,2\tilde{p}_i\tilde{K}}\,\tilde{p}_i^\mu-k_{\perp,12}^\mu\;,
        &
        v_{12}&=v_{1}+v_{2}\;,\quad k_{\perp,12}=k_{\perp,1}+k_{\perp,2}\;.
    \end{split}
\end{equation}
We find the following identities for the magnitude of the transverse momenta
\begin{equation}\label{eq:def_kt2_identified_nnlo}
  \begin{split}
  {\rm k}_{\perp,1/2}^2=&\;2\tilde{p}_i\tilde{K}\,v_{1/2}\,
  \big[\,z_{1/2}-v_{1/2}(1-z+\kappa)\,\big]\;,
  &&\text{where}\qquad
  z_{1/2}=\frac{q_{1/2}N}{\tilde{p}_iN}\\
  {\rm k}_{\perp,12}^2=&\;2\tilde{p}_i\tilde{K}\,v_{12}\,
  \big[\,(1-z+\kappa)(1-v_{12})-\kappa\,\big]
  -(1-v_{12})\,q_{12}^2\;,
  &&\text{where}\qquad
  \;\,q_{12}^2=-\,v_1v_2\bigg(\frac{k_{\perp,1}}{v_1}-\frac{k_{\perp,2}}{v_2}\bigg)^2\;.
  \end{split}
\end{equation}
We can insert this relation into Eq.~\eqref{eq:def_pi_K_nnlo} to make
both collinear safety and overall four-momentum conservation manifest.
\begin{equation}\label{eq:pi_K_using_kt2_nnlo}
    \begin{split}
        q_{1/2}^\mu&=z_{1/2}\,\tilde{p}_i^\mu
        +v_{1/2}\big(\tilde{K}^\mu-(1-z+2\kappa)\,\tilde{p}_i^\mu\big)
        +k_{\perp,1/2}^\mu\;,\\
        K^\mu&=\tilde{K}^\mu-\sigma_{12}\,\tilde{p}_i^\mu
        -v_{12}\big(\tilde{K}^\mu-(1-z+2\kappa)\,\tilde{p}_i^\mu\big)
        -k_{\perp,12}^\mu\;,
    \end{split}
\end{equation}
where $\sigma_{12}=q_{12}^2/(2\tilde{p}_i\tilde{K})$.
The factorized expression for the four-particle phase space reads
(see Ref.~\cite{Assi:2023rbu} for details on the notation)
\begin{equation}\label{eq:emission_phase_space_fi_rad_nnlo}
    {\rm d}\Phi_{4}(-K;p_i,q_1,q_2,Q)
    ={\rm d}\Phi_{2}(-K;q_2,-M)\,
    \frac{{\rm d}M^2}{2\pi}\,
    {\rm d}\Phi_{2}(-M;q_1,-N)
    \frac{{\rm d}N^2}{2\pi}\,
    {\rm d}\Phi_{2}(-N;p_i,Q)\;,
\end{equation}
where $Q^\mu=\sum_{l\neq i,j}p_l^\mu$ is the sum of all final-state
momenta except the collinear triple $p_i^\mu$, $q_1^\mu$ and $q_2^\mu$,
and $M^\mu=N^\mu-q_1^\mu$. The differential two-body phase-space element
for the production of $N^\mu$ and $q_1^\mu$ is given by
\begin{equation}\label{eq:pj_production_ps_ff_1}
  \begin{split}
    {\rm d}{\Phi}_2(-M;q_1,-N)=&\;\frac{(4\pi^2)^{\eps}}{16\pi^2}\,
    \frac{(q_1N)^{1-2\eps}}{(N^2)^{1-\eps}}\,{\rm d}\Omega_{1,N}^{2-2\eps}
    =\frac{(2\tilde{p}_{i}\tilde{K})^{-\eps}}{(16\pi^2)^{1-\eps}}\,
    \frac{z_1^{1-2\eps}}{2(1-z+\kappa)^{1-\eps}}\,{\rm d}\Omega_{1,N}^{2-2\eps}\;,
  \end{split}
\end{equation}
while the two-body phase-space element for the production of
$M^\mu$ and $q_2^\mu$ in the frame of $N^\mu$ reads
\begin{equation}\label{eq:pj_production_ps_ff_2}
  \begin{split}
    {\rm d}{\Phi}_2(-K;q_2,-M)=&\;\frac{(4\pi^2)^{\eps}}{16\pi^2}\,\frac{q_2N}{q_2K}\,
    \frac{(q_2N)^{1-2\eps}}{(N^2)^{1-\eps}}\,{\rm d}\Omega_{2,N}^{2-2\eps}
    =\frac{(2\tilde{p}_{i}\tilde{K})^{-\eps}}{(16\pi^2)^{1-\eps}}\,
    \frac{z_2}{z_2-\sigma_{12}}\,
    \frac{z_2^{1-2\eps}}{2(1-z+\kappa)^{1-\eps}}\,{\rm d}\Omega_{2,N}^{2-2\eps}\;.
  \end{split}
\end{equation}
Finally, we combine Eqs.~\eqref{eq:emission_phase_space_fi_rad_nnlo}, 
\eqref{eq:pj_production_ps_ff_1} and~\eqref{eq:pj_production_ps_ff_2}
to obtain the two-emission phase space element
\begin{equation}\label{eq:emission_phase_space_if_ml_nnlo}
  \begin{split}
  {\rm d}\Phi_{+2}^{(a)}(-\tilde{K},\tilde{p}_{i};q_1,q_2)
  =&\;\bigg(\frac{2\tilde{p}_{i}\tilde{K}}{16\pi^2}\bigg)^{2-2\eps}\,
  \frac{\big(z_1z_2z\big)^{1-2\eps}}{
    (1-z+\kappa)^{2-2\eps}}\,
  \frac{z_2}{z_2-\sigma_{12}}\,
  {\rm d}z\,{\rm d}z_1\,
  \frac{{\rm d}\Omega_{1,N}^{2-2\eps}}{4\pi}\,
  \frac{{\rm d}\Omega_{2,N}^{2-2\eps}}{4\pi}\;.
  \end{split}
\end{equation}
where we have used the fact that $M^2=2\tilde{p}_i\tilde{K}(1-z+\kappa-z_1)$.
In this form, the differential phase space element is not useful for integration,
because $\sigma_{12}$ depends non-trivially on the energy fractions $z_1$ and $z_2$.
We therefore perform a change of variables
\begin{equation}
    v_{1/2}\to\eta_{1/2}=v_{1/2}\,\frac{1-z+\kappa}{z_{1/2}}
    =\frac{(p_iq_{1/2})N^2}{2(p_iN)(q_{1/2}N)}\;.
\end{equation}
Using this definition, we have
\begin{equation}
    \sigma_{12}=\frac{z_1z_2\,\eta_{12}}{1-z+\kappa}\;,
    \qquad\text{where}\qquad
    \eta_{12}=\eta_1+\eta_2-2\eta_1\eta_2
    -2\sqrt{\eta_1(1-\eta_1)\eta_2(1-\eta_2)}\cos\phi_{12}\;.
\end{equation}
We can rewrite this with the help of the identity $1-z=z_1+z_2-\sigma_{12}$,
which allows to express $z_2$ in terms of the integration variables.
Defining $\zeta=z_1/(1-z)$, Eq.~\eqref{eq:emission_phase_space_if_ml_nnlo}
takes the form
\begin{equation}\label{eq:emission_phase_space_if_ml_nnlo_2}
  \begin{split}
  {\rm d}\Phi_{+2}^{(a)}(-\tilde{K},\tilde{p}_{i};q_1,q_2)
  =&\;\bigg(\frac{2\tilde{p}_{i}\tilde{K}}{16\pi^2}\bigg)^{2-2\eps}\,
  \frac{(1-z)^{3-4\eps}\big(z\,\zeta(1-\zeta)\big)^{1-2\eps}}{
    \big((1-z)(1-\zeta\,\eta_{12})+\kappa\big)^{2-2\eps}}\,
  {\rm d}z\,{\rm d}\zeta\,
  \frac{{\rm d}\Omega_{1,N}^{2-2\eps}}{4\pi}\,
  \frac{{\rm d}\Omega_{2,N}^{2-2\eps}}{4\pi}\;.
  \end{split}
\end{equation}
Note that the double-soft limit corresponds to $z\to 1$, while the
single-soft limit is reached for $\zeta\to 0$ or $\zeta\to 1$. 
Equation~\eqref{eq:emission_phase_space_if_ml_nnlo_2} can be used to
compute the 2-emission phase-space volume in the back-to-back configuration,
$\kappa=-1$, which gives, in agreement with Eq.~(4.14) of
Ref.~\cite{Gehrmann-DeRidder:2003pne},
\begin{equation}\label{eq:ps_volume_4}
  \Phi_{+2}(-\tilde{K})=\frac{(4\pi)^{2\eps}}{(16\pi^2)^2}
  \frac{\Gamma(2-2\eps)\Gamma(1-\eps)^3}{\Gamma(4-4\eps)\Gamma(3-3\eps)}\,
  \big(\tilde{K}^2\big)^{2-2\eps}\;.
\end{equation}
Changing the integration variables in Eq.~\eqref{eq:emission_phase_space_if_ml_nnlo}
from $z$ and $z_1$ to $\xi_1$ and $\xi_2$, where
\begin{equation}\label{eq:def_xi_nnlo}
    \xi_{1/2}=\frac{\kappa\,z_{1/2}}{1-z+\kappa-z_{1/2}}\;,
\end{equation}
one can derive a symmetric form of the phase-space element,
which is better suited for numerical integration:
\begin{equation}\label{eq:emission_phase_space_if_ml_nnlo_3}
  \begin{split}
  {\rm d}\Phi_{+2}^{(a)}(-\tilde{K},\tilde{p}_{i};q_1,q_2)
  =&\;\bigg(\frac{\tilde{K}^2}{16\pi^2}\bigg)^{2-2\eps}\,
  \frac{\big(z\,\xi_1\xi_2\big)^{1-2\eps}\,{\rm d}\xi_1\,{\rm d}\xi_2}{
    \big(1-\xi_1\xi_2(1-\eta_{12})/\kappa^2\big)^{3-2\eps}}\,
  \frac{{\rm d}\Omega_{1,N}^{2-2\eps}}{4\pi}\,
  \frac{{\rm d}\Omega_{2,N}^{2-2\eps}}{4\pi}\;,
  \end{split}
\end{equation}
where
\begin{equation}\label{eq:def_z_nnlo_ps}
  z=1+\kappa-\frac{\kappa(\xi_1+\kappa)(\xi_2+\kappa)}{
    \kappa^2-\xi_1\xi_2(1-\eta_{12})}\;.
\end{equation}
The physically allowed phase space in this parametrization
is given by $0\le\xi_{1,2}\le-\kappa$ and $0\le\eta_{12}
  \le1+\kappa/(1+\kappa)(1-\kappa(1-\xi_1-\xi_2)/(\xi_1\xi_2))$,
with $0\le\eta_{12}\le1$ in the back-to-back case, $\kappa=-1$.
One can now use the sector decomposition developed in
Refs.~\cite{Czakon:2010td,Czakon:2011ve}, which is based
on the following transformation of one of the azimuthal
angle integrals:
\begin{equation}
  (\sin\phi_{12})^{-2\eps}\,{\rm d}\phi_{12}=
  \frac{\eta_3^{1-2\eps}}{|\eta_1-\eta_2|^{1-2\eps}}\,
  (4\chi(1-\chi))^{-\frac{1}{2}-\eps}\,2\,{\rm d}\chi\;,
\end{equation}
where
\begin{equation}
  \eta_3=\frac{(\eta_1-\eta_2)^2}{\eta_{12}}
  \qquad\text{and}\qquad
  \chi=\frac{1}{2}\frac{(1-\cos(\theta_1-\theta_2))(1+\cos\phi_{12})}{
    1-\cos(\theta_1-\theta_2)+(1-\cos\phi_{12})\sin\theta_1\sin\theta_2}\;.
\end{equation}
The mapping of the sector integration variables, which we denote by
$\hat{\xi}_1$, $\hat{\xi}_2$, $\hat{\eta}_1$, and $\hat{\eta}_2$,
to the individual phase-space variables is reproduced in
Tab.~\ref{tab:mapping_sector_variables} for completeness.
\begin{table}
    \setlength{\tabcolsep}{10pt}
    \renewcommand{\arraystretch}{1.75}
    \begin{tabular}{c|c|c|c|c|c|c|c}
    Type & \multicolumn{5}{c|}{Triple collinear} & \multicolumn{2}{c}{Double collinear} \\\hline
    Sector & $S^I_1$ & $S^I_2$ & $S^I_3$ & $S^I_4$ & $S^I_5$ & $S^{II}_1$ & $S^{II}_2$ \\\hline
    $\eta_1$ & $\hat{\eta}_1$ & $\frac{1}{2}\hat{\eta}_1\hat{\eta}_2$ &
      $\frac{1}{2}\hat{\eta}_1\hat{\eta}_2\xi_2$ & $\hat{\eta}_1$ &
      $\frac{1}{2}(2-\hat{\eta}_1)\hat{\eta}_2$ & $\hat{\eta}_1$ & $\hat{\eta}_1\hat{\xi}_2$ \\\hline
    $\eta_2$ & $\frac{1}{2}\hat{\eta}_1\hat{\eta}_2$ & $\hat{\eta}_2$ &
      $\hat{\eta}_2$ & $\frac{1}{2}\hat{\eta}_1(2-\hat{\eta}_2)$ &
      $\hat{\eta}_2$ & $1-\hat{\eta}_2$ & $1-\hat{\eta}_2$ \\\hline
    $\xi_1$ & $\hat{\xi}_1$ & $\hat{\xi}_1$ & $\hat{\xi}_1$ &
      $\hat{\xi}_1$ & $\hat{\xi}_1$ & $\hat{\xi}_1$ & $\hat{\xi}_1$ \\\hline
    $\xi_2$ & $\hat{\xi}_1\hat{\xi}_2$ & $\hat{\eta_1} \hat{\xi}_1\hat{\xi}_2$ &
      $\hat{\xi}_1\hat{\xi}_2$ & $\hat{\xi}_1\hat{\xi}_2$ &
      $\hat{\xi}_1\hat{\xi}_2$ & $\hat{\eta}_1 \hat{\xi}_1\hat{\xi}_2$ & $\hat{\xi}_1\hat{\xi}_2$ \\
    \end{tabular}
    \caption{Mapping of sector integration variables $\hat{\xi}_1$,
    $\hat{\xi}_2$, $\hat{\eta}_1$, $\hat{\eta}_2$ to the phase-space
    variables in Eq.~\eqref{eq:emission_phase_space_if_ml_nnlo_3},
    following the algorithm in Refs.~\cite{
      Czakon:2010td,Czakon:2011ve}\label{tab:mapping_sector_variables}}
\end{table}

Finally we discuss a phase-space parametrization in iterated
NLO kinematics with different radiators. It is obtained by
applying Eq.~\eqref{eq:emission_phase_space_if_ml} twice.
If the particles with momenta $q_1$ and $q_2$ are emitted sequentially,
and the longitudinal recoil is taken by the momenta $p_i$ and $p_k$,
respectively, we find
\begin{equation}\label{eq:iterated_nlo_kinematics_2}
  {\rm d}\Phi_{+2}(-\tilde{K},\tilde{p}_{i},\tilde{p}_{k};q_1,q_2)
  =\bigg(\frac{2\tilde{p}_i\tilde{K}}{16\pi^2}\bigg)^{2-2\eps}\,
  \frac{\big(z_1(1-z_1)z_2(1-z_2)\big)^{1-2\eps}}{
    (z_1+\kappa)^{1-\eps}(z_2+\kappa)^{1-\eps}}\,
  \mathcal{J}_{i,1;k,2}^{\,1-\eps}\,{\rm d}z_1\,{\rm d}z_2\,
  \frac{{\rm d}\Omega_{1,N}^{2-2\eps}}{4\pi}
  \frac{{\rm d}\Omega_{2,M}^{2-2\eps}}{4\pi}\,\;,
\end{equation}
where $N^\mu=\tilde{K}^\mu+z_1\tilde{p}_i^\mu$,
$M^\mu=N^\mu-q_1^\mu+z_2\tilde{p}_k^\mu$, and
\begin{equation}\label{eq:iterated_nlo_kinematics_jac}
  \mathcal{J}_{i,1;k,2}=
  \frac{(z_2+\kappa)\big[\,2\tilde{p}_{k}
    (\tilde{K}+z_1\tilde{p}_i-q_1)\,\big]^2}{
    2\tilde{p}_i\tilde{K}\big[\,\tilde{K}^2+z_2\,
    2\tilde{p}_k(\tilde{K}+z_1\tilde{p}_i-q_1)\big]}\;.
\end{equation}
In the back-to-back case, this reduces to
\begin{equation}\label{eq:iterated_nlo_kinematics_jac_b2b}
  \mathcal{J}_{i,1;k,2}=\bigg(1+\frac{2p_iq_1}{2\tilde{p}_{i}\tilde{K}}\bigg)^2\,
    \bigg[1-\frac{z_2}{1-z_2}\frac{2p_iq_1}{2\tilde{p}_i\tilde{K}}\bigg]^{-1}\;.
\end{equation}

\bibliographystyle{apsrev4-1}
\bibliography{main}
\end{document}